\documentclass{aa}  

\usepackage{comment}
\usepackage{graphicx}
\usepackage[dvipsnames]{xcolor}
\usepackage{caption}
\usepackage{subcaption}

\usepackage{pdflscape} 
\usepackage{txfonts}
\usepackage{hyperref} 
\hypersetup{colorlinks=true, linktocpage=true, pdfstartpage=3,
            pdfstartview=FitV, breaklinks=true, pdfpagemode=UseNone,
            pageanchor=true, pdfpagemode=UseOutlines, plainpages=false,
            bookmarksnumbered, bookmarksopen=true, bookmarksopenlevel=1,
            hypertexnames=true, urlcolor=RoyalBlue,
            linkcolor=RoyalBlue, citecolor=RoyalBlue}

\usepackage[utf8]{inputenc}
\usepackage[T1]{fontenc}

\begin{document}

   \title{Hubble Asteroid Hunter}

   \subtitle{I. Identifying asteroid trails in Hubble Space Telescope images}

   \author{Sandor Kruk\inst{1,2}\thanks{e-mail: kruksandor@gmail.com},
          Pablo Garc\'ia Mart\'in \inst{3}, Marcel Popescu \inst{4}, Bruno Mer\'in  \inst{5},  Max Mahlke \inst{6}, Beno\^it Carry\inst{6}, Ross Thomson \inst{7}, Samet Karada\u{g}\inst{8}, Javier Dur\'an\inst{9}, Elena Racero\inst{10}, Fabrizio Giordano\inst{10}, Deborah Baines\inst{11}, Guido de Marchi\inst{1}, Ren\'e Laureijs\inst{1}
          }

   \institute{Max-Planck-Institut für extraterrestrische Physik (MPE), Giessenbachstrasse 1, D-85748 Garching bei München, Germany
   \and
       European Space Agency (ESA), European Space Research and Technology Centre (ESTEC), Keplerlaan 1, 2201 AZ Noordwijk, The Netherlands
   \and
  Departamento de F\'{\i}sica Te\'orica, Universidad Aut\'onoma de Madrid, Madrid 28049, Spain
  \and
  Astronomical Institute of the Romanian Academy, 5 Cuţitul de Argint, 040557 Bucharest, Romania
   \and   
  European Space Agency (ESA), European Space Astronomy Centre (ESAC), Camino Bajo del Castillo s/n, 28692 Villanueva de la Cañada, Madrid, Spain
  \and
  Universit\'e C\^ote d'Azur, Observatoire de la C\^ote d'Azur, CNRS, Laboratoire Lagrange, France
   \and
  Google Cloud, 6425 Penn Ave, Pittsburgh, PA 15206, United States
  \and
  Google, Claude Debussylaan 34, 1082 MD Amsterdam, The Netherlands
  \and
  RHEA for European Space Agency (ESA), European Space Astronomy Centre (ESAC), Camino Bajo del Castillo s/n, 28692 Villanueva de la Cañada, Madrid, Spain
  \and
  SERCO for European Space Agency (ESA), European Space Astronomy Centre (ESAC), Camino Bajo del Castillo s/n, 28692 Villanueva de la Cañada, Madrid, Spain
  \and
  QUASAR SCIENCE RESOURCES for European Space Agency (ESA), European Space Astronomy Centre (ESAC), Camino Bajo del Castillo s/n, 28692 Villanueva de la Cañada, Madrid, Spain}

   \date{Received December 24, 2021; accepted January 25, 2022}

\titlerunning{Hubble Asteroid Hunter: Identifying asteroid trails in Hubble Space Telescope images}
\authorrunning{Kruk et al.}
% \abstract{}{}{}{}{} 
% 5 {} token are mandatory
 
  \abstract
  {Large and publicly available astronomical archives open up new possibilities to search for and study Solar System objects.  However, advanced techniques are required to deal with the large amounts of data. These unbiased surveys can be used to constrain the size distribution of minor bodies, which represents a piece of the puzzle for the formation models of the Solar System.}
   {We aim to identify asteroids in archival images from the ESA Hubble Space Telescope (\textit{HST}) Science data archive using data mining.}  
   {We developed a citizen science project on the Zooniverse platform, Hubble Asteroid Hunter (\url{www.asteroidhunter.org}), and have asked members of the public to identify asteroid trails in archival \textit{HST} images. We used the labels provided by the volunteers to train an automated deep learning model built with Google Cloud AutoML Vision to explore the entire \textit{HST} archive to detect asteroids crossing the field-of-view.}
   {We report the detection of 1\,701 new asteroid trails identified in archival \textit{HST} data via our citizen science project and the subsequent machine learning exploration of the ESA \textit{HST} science data archive. We detect asteroids to a magnitude of 24.5, which are statistically fainter than the populations of asteroids identified from ground-based surveys. The majority of asteroids are distributed near the ecliptic plane, as expected, where we find an approximate density of 80 asteroids per square degree. We matched 670 trails (39\% of the trails found) with 454 known Solar System objects in the Minor Planet Center database; however, no matches were found for 1\,031 (61\%) trails. The unidentified asteroids are faint, on average 1.6 magnitudes fainter than the asteroids we succeeded in identifying. They probably correspond to previously unknown objects.}
   {Citizen science and machine learning are very useful techniques for the systematic search for Solar System objects in existing astronomy science data archives. This work describes a method for finding new asteroids in astronomical archives that span decades; it could be effectively applied to other datasets, increasing the overall sample of well-characterised small bodies in the Solar System and refining their ephemerides.}


   \maketitle
%
%-------------------------------------------------------------------

\section{Introduction}

Solar System objects (SSOs), asteroids and comets, represent the remnants of the planetesimals that once formed the planets. Understanding SSOs provides key constraints on the evolution of the Solar System and on the evolution of other planetary systems.

In recent years there has been an exponential increase in asteroid discoveries thanks to robotic telescopic surveys and powerful detection algorithms. As of 10 December 2021, the orbits of more than 1.1 million SSOs are listed by the Minor Planet Center (MPC)\footnote{\url{https://www.minorplanetcenter.net/}}, the main worldwide repository for the receipt and distribution of positional measurements of SSOs. Most of these discoveries have been obtained with 1-2 m class telescopes, such as Pan-STARRS\footnote{\url{http://pswww.ifa.hawaii.edu/pswww/}} or the Catalina Sky Survey\footnote{\url{https://catalina.lpl.arizona.edu/}}, as part of survey efforts dedicated to the observation and characterisation of objects on near-Earth orbits. The magnitude limit for individual exposures for such telescopes is approximately 23 mag or brighter, depending on the filter 
(Kaiser et al. \citeyear{Kaiser2010}, Drake et al. \citeyear{Drake2014a} and \citeyear{Drake2017}). 

Identifying faint asteroids is important as their distribution at the smallest sizes is poorly understood. \citet{2015aste.book..701B} argue the wavy shape of the size distribution curve is a byproduct of comminution as one goes to smaller sizes, with its shape a fossil-like remnant of a violent early epoch.  Thus, in order to test the various collisional models, the key data are represented by the smallest bodies, which are less well known due to their faint magnitudes and can only be discovered using large telescopes or from space.  Moreover, small bodies are those most affected by non-gravitational effects, such as the Yarkovsky or the Yarkovsky–O'Keefe–Radzievskii–Paddack (YORP) effects \citep{Bottke2006, Vokrouhlicky2015}, and their long-term evolution is erratic \citep{2015Icar..247..191B}.

While space-based observatories generally have smaller fields-of-view (FoVs) compared to ground-based ones, they offer a unique opportunity to study SSOs. For example, the Hubble Space Telescope (\textit{HST}) can observe a magnitude of up to V=27 for point sources (for a 1 hour exposure; \citealt{Ryon2021}) and, with an image resolution of 0.09 arcsec, is able to resolve small displacements. With images taken over more than three decades,  the \textit{HST} archives provide one of the longest time baselines available for studying SSOs. Although they are not the target of \textit{HST} observations, the serendipitous detection of trails can enable an unbiased study of asteroid occurrences and properties as well as the recovery of interesting objects. An example is the serendipitous observation of (16) Psyche identified in Herschel images \citep{Racero2022}.

In its 31 years of observations, \textit{HST} has produced a rich archive of hundreds of thousands of targeted observations of nebulae, galaxies,  clusters of galaxies, and gravitational lenses. Occasionally, closer objects such as asteroids cross the telescope's FoV while the targets are being observed, leaving trails in the images. More than two decades ago, \citet{Evans1998} investigated the presence of asteroid trails in archival \textit{HST} images taken with its original camera, the Wide Field and Planetary Camera 2 (WFPC2). Even though two new instruments have been installed to replace the WFPC2, the Advanced Camera for Surveys (ACS) and Wide Field Camera 3 (WFC3), the serendipitous presence of asteroids in the \textit{HST} images has not received more attention.

In 2019, on International Asteroid Day, we launched the Hubble Asteroid Hunter\footnote{\url{www.asteroidhunter.org}} citizen science project on the Zooniverse platform, with the goals of visually identifying asteroids in archival images from the European Space Agency \textit{HST}  (eHST)\footnote{\url{http://hst.esac.esa.int/ehst}} archive and studying their properties. 

With a low rate of trails in the images, visually detecting asteroid is time-consuming. Machine learning has shown great potential as a method for classifying large amounts of data rapidly and has been applied to tackle various problems in astronomy: galaxy classification \citep{Dieleman2015,Huertas-Company2015,Walmsley2020,Walmsley2022}, detection of strong gravitational lenses \citep{Canameras2021}, and estimating photometric redshifts \citep{Pasquet2019}. Deep learning methods, and in particular convolutional neural networks (CNNs), have recently been applied to automate the detection of SSOs (e.g. \citealt{Lieu2019,Duev2019}). 

In this study, we investigate the use of an automated machine learning (AutoML) algorithm used for both the detection and classification of trails in images from \textit{HST}, both for trails produced by SSOs and for objects that can be confused with SSOs: artificial satellites, artefacts (cosmic rays), and extragalactic sources (arcs of strong gravitational lenses).

In Sect. \ref{datasection} we discuss the \textit{HST} data used in this study. Section \ref{methodsection} describes the methods used for the work, citizen science, and deep learning. In Sect. \ref{resultssection} we present the results, and in Sect. \ref{discussionsection} we discuss their implications. The final conclusions of the paper are given in Sect. \ref{conclusionssection}.
 
%--------------------------------------------------------------------
\section{Hubble observations and data used}
\label{datasection}

This project uses images from the \textit{HST} Advanced Camera for Surveys Wide Field Camera (ACS/WFC) and the Wide Field Camera 3 Ultraviolet and Visible Channel (WFC3/UVIS). These are the two \textit{HST} instruments with the largest FoVs, which therefore have the highest chance of containing asteroid trails. We refer the analysis of near-infrared images from WFC3/IR channel to a later work. SSOs, with a spectral energy distribution approximately of a G2 star, are fainter at 1-2 $\mu$m. Additionally, the smaller FoV, different resolution, and various artefacts present in the near-IR images make it more difficult to identify asteroid trails in these images and directly compare them with the optical images.

\begin{figure*}
   \centering
   \includegraphics[width=\textwidth]{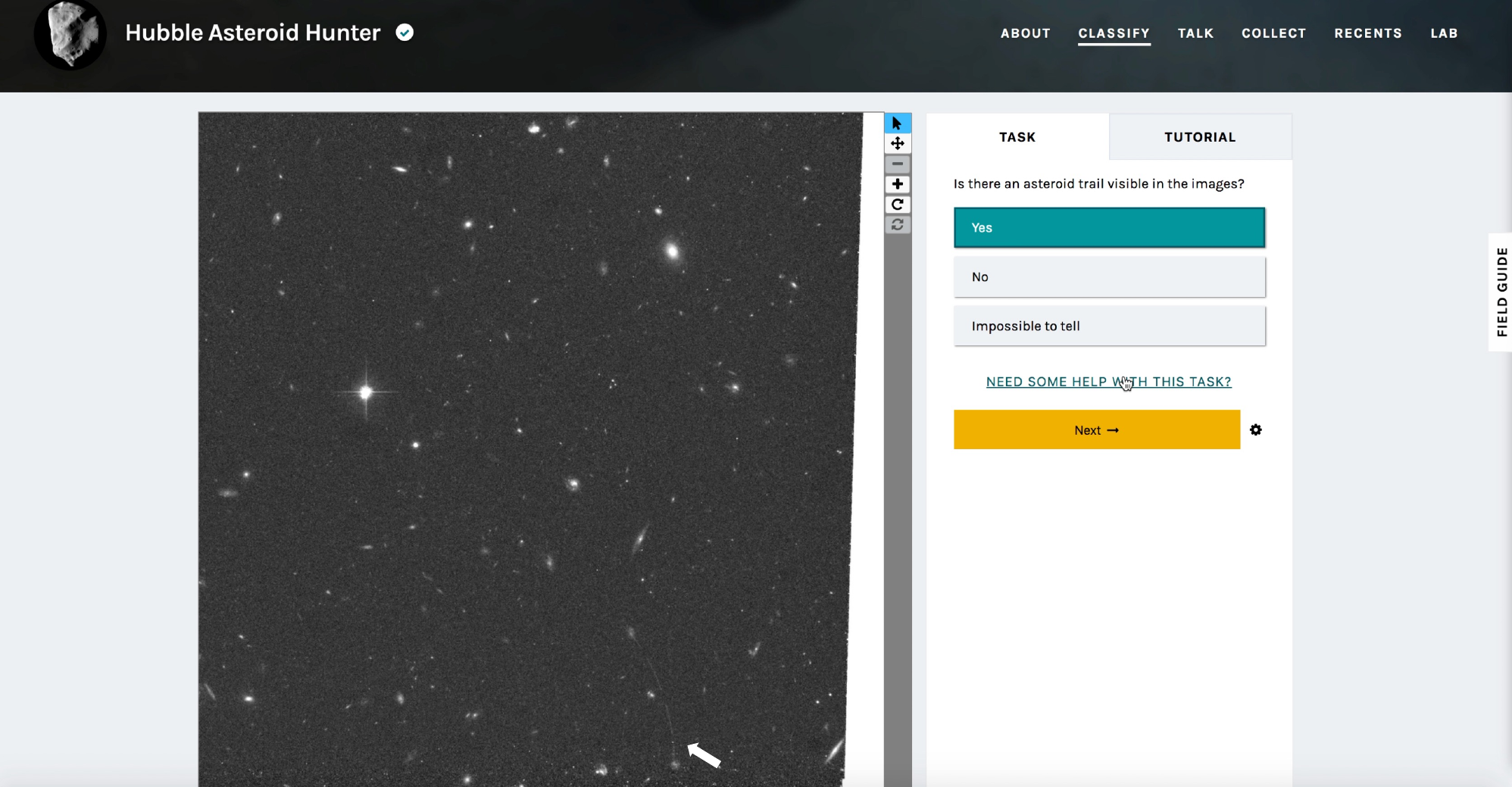}
   \caption{Image showing the classification interface of the Hubble Asteroid Hunter (\url{www.asteroidhunter.org}) citizen science project. Individual users are asked to inspect \textit{HST} image quadrants for asteroids and mark the beginning and end point of trails in the images. Markings from ten users are aggregated into a final classification. The asteroid trail is visible as a `C' shape in the bottom right of the exposure. There are gaps in the trail as the displayed cutout is an \textit{HST} composite exposure. The white arrow shows the position of the trail.}
    \label{interface}%
\end{figure*}

The images were processed using the standard pipeline calibration settings and were taken directly from eHST. Individual dithered observations are aligned and processed by DrizzlePac\footnote{\url{https://hst-docs.stsci.edu/drizzpac}} \citep{Gonzaga2012} for geometric distortion corrections and combined to remove cosmic rays. However, this correction technique is not suitable for moving targets, such as asteroids. The DrizzlePac algorithm flags the trails produced by moving targets as cosmic rays and attempts to remove them. These can appear as residuals in the composite images, which remain visible in the archival images. Identifying trails on individual, \textit{HST} short exposures is challenging due to the presence of many cosmic rays, which may induce a high number of false positives. Moreover, combining multiple images together, as in the case of \textit{HST} composite images, makes the trails appear longer so easier to detect. One example of an asteroid trail in the combined \textit{HST} observations is shown in Fig. \ref{interface}. Asteroids often appear as curved streaks due to the parallax induced by the orbital motion of \textit{HST} during the exposures and the moving asteroid. Depending on distance to the asteroids, the pointing of \textit{HST,} and the relative motion of the asteroid compared to \textit{HST,} they appear as having a gentle curvature, a `C' shape, or a more extreme `S' shape. 

In this project, we analysed archival Hubble images taken between 30 April 2002 (when the ACS camera was installed) and 14 March 2021 for ACS/WFC and 24 June 2009 (when the WFC3 obtained first light) and 14 March 2021. Therefore, the analysis is complete with data taken and publicly available in the \textit{HST} archives up to 14 March 2021. We note that observations based on general observer (GO) proposals are available in the \textit{HST} archive one year after they were taken; therefore, the last GO observation analysed were taken on 14 March 2020. \textit{HST} Snapshot observations are available in the archive immediately after they were acquired and they were analysed up to this date.

All the composite \textit{HST} images were selected from eHST based on the following criteria: an exposure time greater than 100 seconds and a FoV greater than 0.044$^{\circ}$ (to exclude sub-frames). We excluded the grism spectral images, as the spatially extended spectral `wings' can be confused with trails, and calibration images. In the case of the WFC3/UVIS images, we excluded images for which the targets were `Dark' frames. 

We used a total of 37\,323 \textit{HST} composite images in \textsc{PNG} format available from the eHST archive (the exact same as available in the Hubble Legacy Archive database). In order to improve the detection of short trails, and the display of the images on the project website, we split the \textsc{PNG} images equally into four equal quadrants. Each quadrant is approximately 1050 pixels on a side and covers a sky area of 101\arcsec $\times$ 101\arcsec (for ACS) and 80\arcsec $\times$ 80\arcsec (for WFC3/UVIS). We refer to these quadrants as `cutouts'. The total number of images and cutouts used in this study is shown in Table \ref{HSTdata}. 

\begin{table}[]
    \centering
    \begin{tabular}{ | c| c| c |  c |} 
    \hline
    Instrument & Field-of-view & No. images & No. cutouts \\ 
    \hline
    ACS/WFC & 202\arcsec $\times$ 202\arcsec  & 24\,731 & 98\,924 \\ 
    WFC3/UVIS & 160\arcsec $\times$ 160\arcsec & 12\,592 & 50\,368 \\
    \hline
    \end{tabular}
    \caption{Number of archival \textit{HST} composite images and cutouts analysed in this paper.}
    \label{HSTdata}
\end{table}

\section{Asteroid identification methods}
\label{methodsection}

We employed a novel method for identifying asteroid trails in \textit{HST} images, combining citizen science and deep learning. 

\subsection{Hubble Asteroid Hunter citizen science project}

The Hubble Asteroid Hunter citizen science project was built using the Zooniverse Panoptes Project Builder and launched on 21 June 2019, ahead of International Asteroid Day, and finished a year later, in August 2020. It attracted 11\,482 volunteers who provided nearly 2 million classifications for the \textit{HST} cutouts. 

A screenshot of the classification interface is shown in Fig. \ref{interface}. The \textit{HST} cutouts are displayed together with a simple question regarding the presence of asteroids in the images: whether or not there is an asteroid trail visible in the images. To familiarise the volunteers with the appearance of asteroid trails in \textit{HST} images, we set up a second workflow (`Training') in which users could choose to classify 20 \textit{HST} cutouts, half of which contained pre-selected asteroid trails. Finally, after classifying an image, volunteers can comment on it in the forum of the project, Talk\footnote{\url{https://www.zooniverse.org/projects/sandorkruk/hubble-asteroid-hunter/talk}}.

\subsubsection{Workflow}

\begin{figure}
   \centering
   \includegraphics[width=\columnwidth]{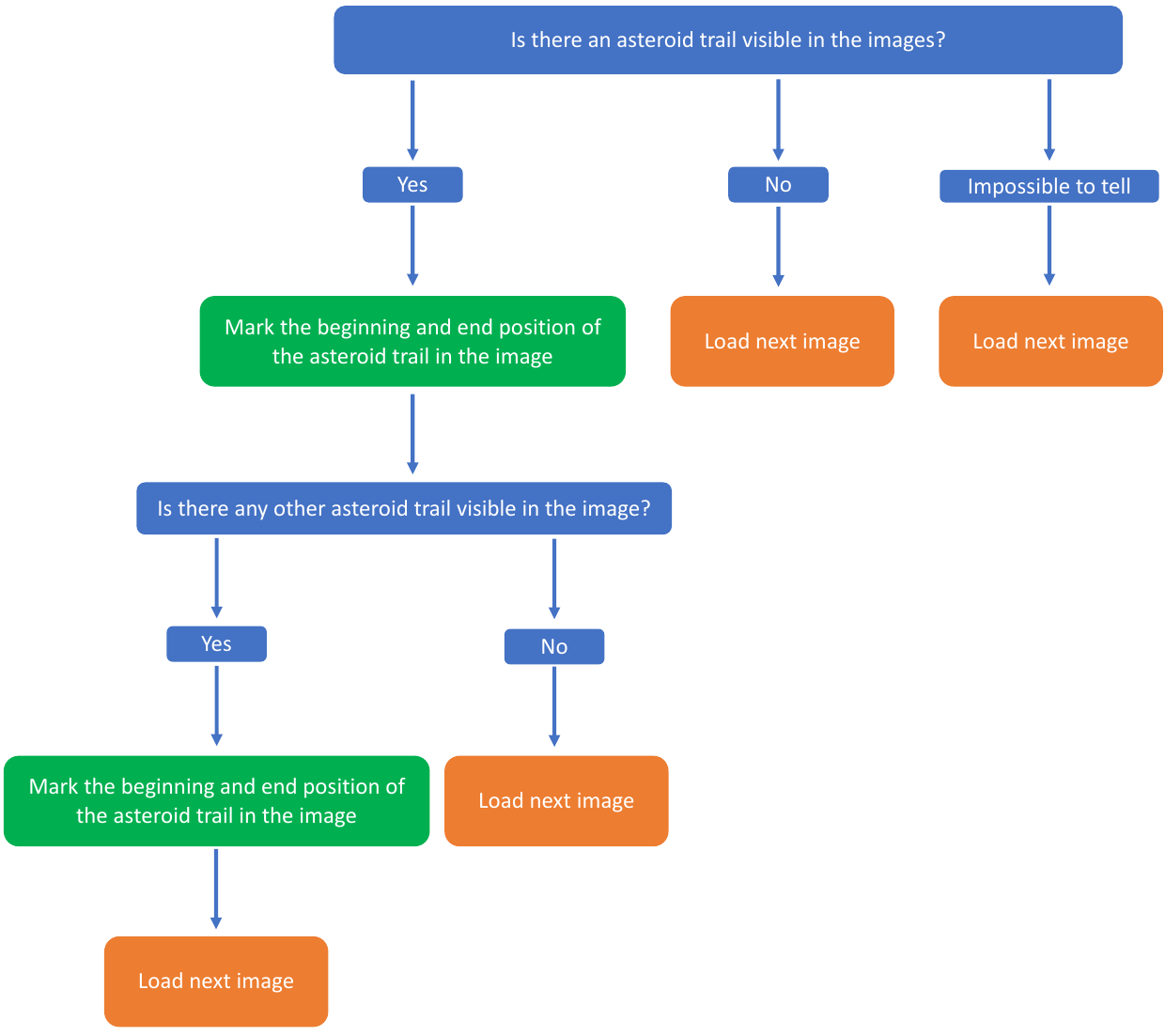}
   \caption{Main workflow of the Hubble Asteroid Hunter citizen science project.}
    \label{workflow}%
\end{figure}

The project consisted of a main workflow (visible in Fig. \ref{workflow}), where we asked the volunteers whether there is an asteroid trail present in the images. If the answer was positive, the users were asked to mark the beginning and end of the asteroid trail with a purposely designed marker tool. If the answer was negative, a new image was presented. In addition, there is an `Impossible to tell' option in case the images were bad and it was impossible to tell whether an asteroids was present. This was the case, for example, for images where \textit{HST} lost tracking and all objects appeared trailing.

The volunteers could mark a second asteroid trail in the cutouts if it was present. We did not include an option to classify more asteroids, as it is very unlikely that there are more than two asteroids in a quadrant. 

\subsubsection{Volunteer training}

At first, volunteers might not be familiar with the appearance of an asteroid trail in the images. To train the citizen scientists, we provided a tutorial for first time users and a separate workflow with a high fraction of asteroid trails; more example images of asteroid trails are shown in the Field Guide as well as in the `Need some help with this task' section. 

The tutorial explained the goals of the project and introduced the task and the typical appearances of trails in the images, the tools available to visualise the images (pan, zoom, and inverting the colours of the image), and objects that could be confused with asteroid trails. 

A second workflow, which had the same structure as the main workflow, was available for training. This workflow contained 20 images, ten (50\%) of which contained asteroid trails of various brightnesses and lengths. Volunteers were prompted to finish classifying these images before proceeding to the main classification task. 

Hubble Asteroid Hunter also provided a static reference set of images for volunteers to consult when in doubt about a particular image. The `Field Guide' contained example images of asteroid trails and some common false positives, such as cosmic rays, `X'-shaped diffraction spikes from stars outside the FoV, edge-on galaxies, satellite trails, and arcs from gravitational lenses in clusters of galaxies. All these features appear as trail-like in the images. Cosmic rays are common, and although the DrizzlePac algorithm is designed to reject them when combining multiple dithered frames, they occasionally appear as subtracted features in the images, which might appear similar to asteroid trails. The difference compared to asteroids is that they are straight, are narrower (not convolved with the point spread function of the instrument), and have a non-uniform brightness. Satellite trails can have a similar width as asteroid trails, but they are straight and cross the entire FoV and are therefore relatively easy to distinguish from asteroids. Gravitational lens arcs, on the other hand, appear curved, similar to asteroid trails, and are abundant in the images of clusters of galaxies, which are frequently observed by \textit{HST}. The arcs, however, are more irregular in shape compared to asteroids and bend around a central object; therefore, they can be distinguished from asteroid trails. The volunteers on the project tagged objects that can be confused with asteroid trails using hashtags (\#satellite, \#cosmic\_ray and \#gravitational\_lens) on the forum of the project. We used these tags in training the automated object detection algorithm described in Sect. \ref{automl}.

\subsubsection{Volunteer classifications}
\label{volunteerclassifications}

The data in the Hubble Asteroid Hunter citizen science project consisted of a subset of all the \textit{HST} ACS/WFC and WFC3/UVIS archival images presented in Table \ref{HSTdata}. Specifically, they consisted of images taken up to 24 April 2020 that had an exposure time greater than 300 seconds. By July 2020, 11\,482 volunteers provided 1\,783\,873 classifications for 144\,559 cutouts. The mean number of classifications per user is 155, with individual counts ranging from 1 classification per user to 85\,333. 

\begin{figure}
   \centering
   \includegraphics[width=\columnwidth]{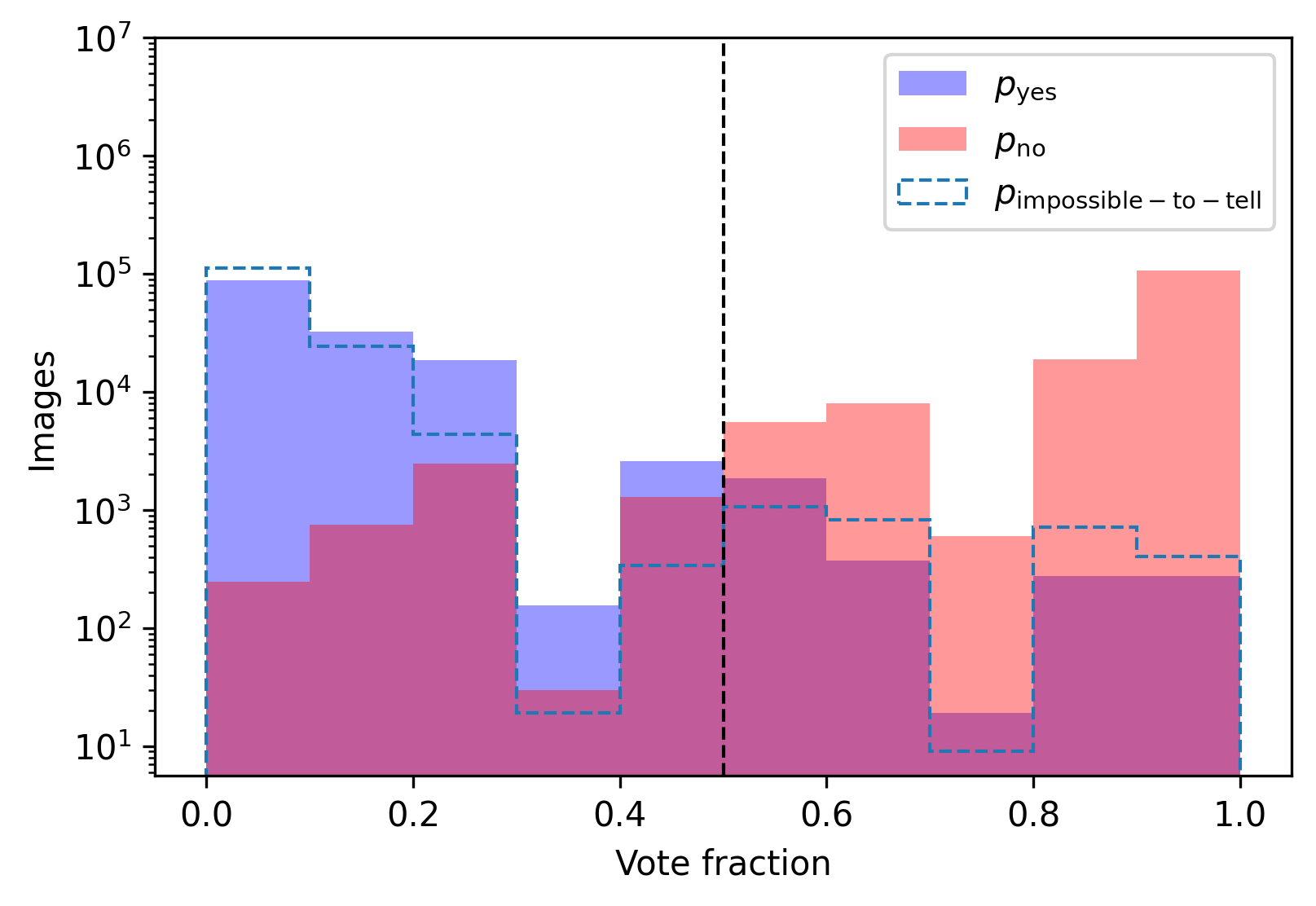}
   \caption{Probability of an \textit{HST} cutout having an asteroid trail, no asteroid trail, or impossible to tell, as voted by ten volunteers on the Hubble Asteroid Hunter project. The distribution is plotted on a log-scale in order to show the low number of positive classifications.}
    \label{classifications}%
\end{figure}

We investigated 144\,559 cutouts, of which 132\,878 were unique  (91\,760 ACS and 41\,118 WFC3/UVIS cutouts) and 11\,681 were uploaded twice to the project. Each cutout was inspected by ten volunteers. For each cutout, we calculated the probability that it contains an asteroid trail ($p_{\mathrm{asteroid}}$), by dividing the number of positive classifications by the total number of classifications. In Fig. \ref{classifications}, we plot the probability of a cutout having an asteroid trail, not having an asteroid, and impossible-to-tell (the three classification options in the project). We took as positive classifications those where the majority of users said it contained an asteroid, $p_{\mathrm{asteroid}}>0.5$, resulting in a total of 1\,488 asteroid trails. Therefore, the volunteers in the project found asteroid trails in only 1\% of the cutouts. We aggregated the markings of the volunteers for the beginning and end point of the trails using a point clustering algorithm, Hierarchical Density-Based Spatial Clustering of Applications with Noise (\textsc{HDBSCAN}; \citealt{Campello2013}). A cluster of points (volunteer clicks) was defined as a group of five or more points clustered together (minimum cluster size = 5) and minimum samples (which provides a measure of how conservative the clustering should be) of 5. In the majority of cases where trails were correctly identified, changing the parameters of \textsc{HDBSCAN} to a larger value had negligible effect on the aggregated positions of the beginning and end point of the trails. The aggregated positions of the trails were used in Sect. \ref{automl} to train the AutoML object detection algorithm as well as to produce smaller cutouts for extracting the trails and performing the photometry. 

\subsubsection{Additional science cases}

The \textit{HST} archives are rich and diverse in the content of astronomical objects. Since it is the first time that humans have visually explored the entire archive in a coordinated fashion, after starting the project we identified several additional science cases. These science cases are related to the identification of rare objects in the archives. Rather than diluting the main science case by adding an additional question to the main workflow, we asked volunteers to tag (with \#gravitational\_lens, \#ring, or \#dwarf) rare objects on the project forum, Talk. Rare objects include strong gravitational lens candidates, collisional and polar rings, and dwarf galaxies. All these objects are subject to further scrutiny by the science team, and they will be catalogued and studied in future publications. 

\begin{figure}
   \centering
   \includegraphics[width=\columnwidth]{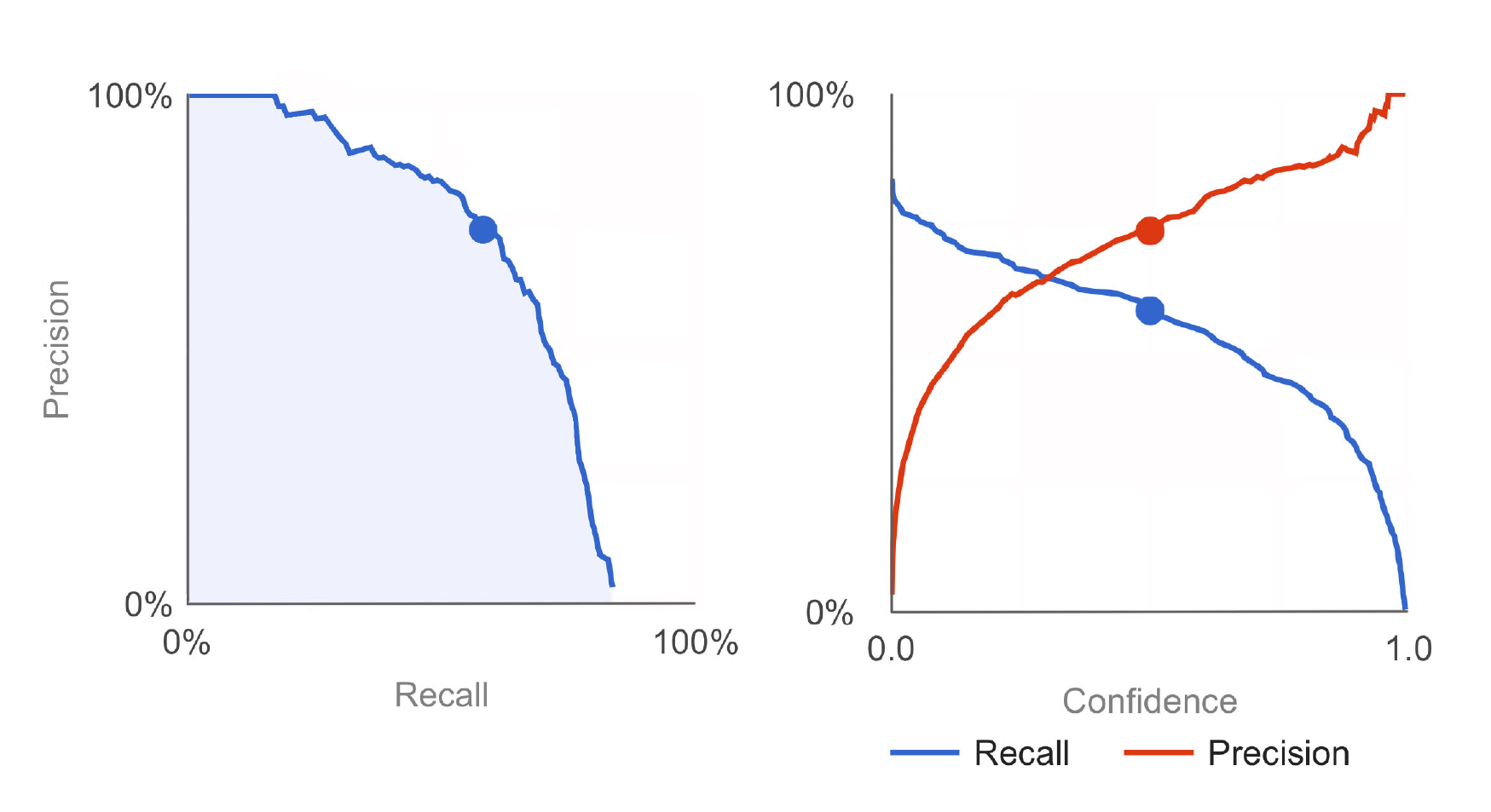}
   \caption{AutoML model performance for asteroid trail detection. The left plot shows the precision vs. recall curve, and the right plot shows both the precision and recall curves as a function of the confidence threshold. In both cases, the circles show the precision and recall values for the 50\% confidence threshold we used for this project.}
    \label{automl_metrics}%
\end{figure}

\begin{figure*}
   \centering
   \includegraphics[width=\textwidth]{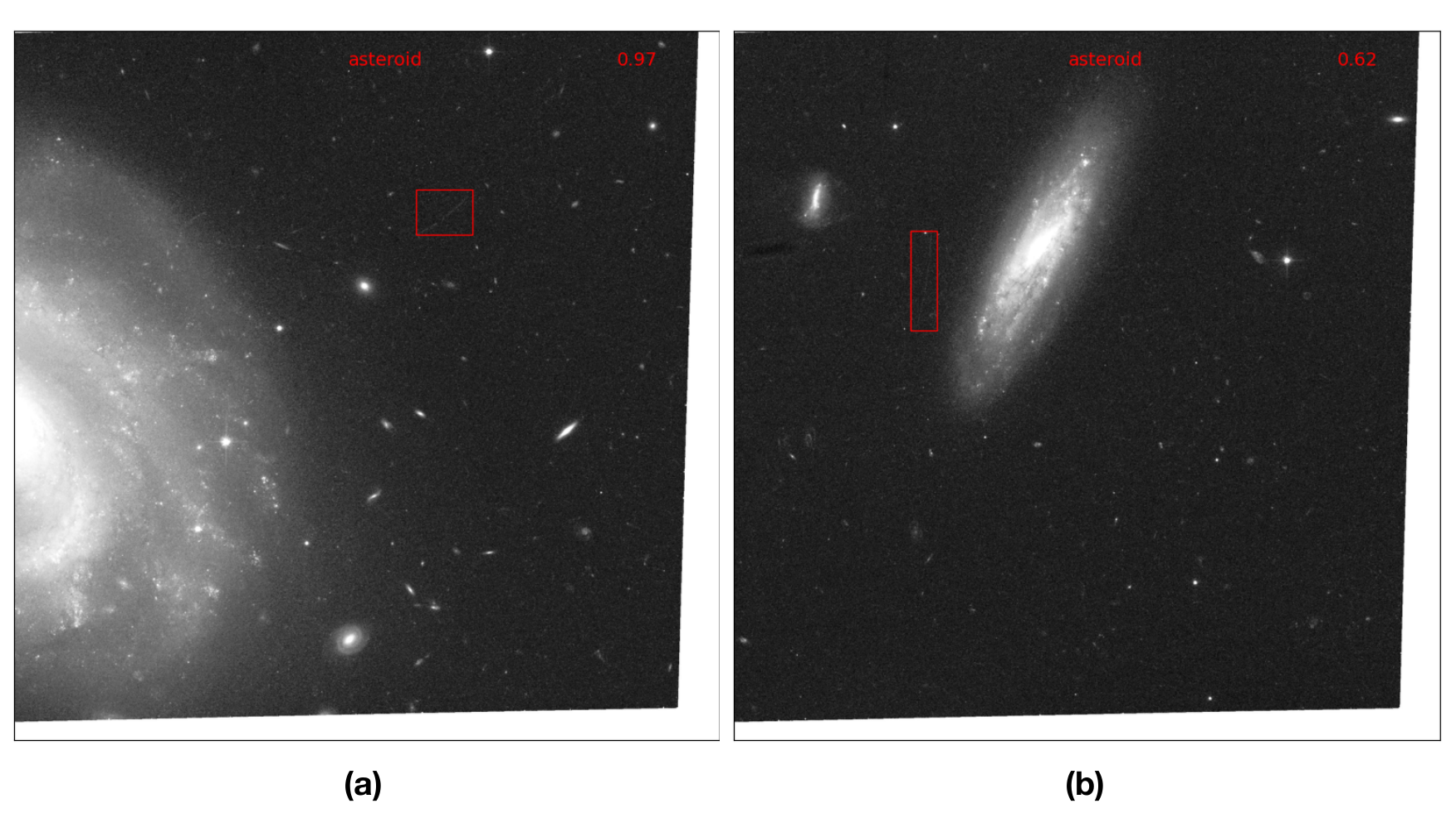}
   \caption{Examples of two asteroids identified by the AutoML deep learning model in the vicinity of galaxies HCG007 (a) and HCG059 (b). Bounding boxes show the position of the trails as detected by the AutoML algorithm. The score in the upper-right part of the images shows the classification confidence of the model.}
    \label{examples_from_automl}%
\end{figure*}

\subsection{AutoML object detection}
\label{automl}

We used the classifications provided by citizen scientists from the Hubble Asteroid Hunter to train an automated multi-object detection algorithm based on CNNs, Google Cloud AutoML Vision\footnote{\url{https://cloud.google.com/vision/automl/object-detection/docs}}. In deep learning, designing the most suitable model is a time-consuming task because of the large space of all possible architectures and parameters. AutoML is an approach for automating the design of machine learning models using a neural architecture search algorithm based on reinforcement learning \citep{Zoph2016}. \citet{Zoph2017} find that AutoML can perform on par with neural networks designed by human experts when testing on the COCO and ImageNet datasets\footnote{\url{https://ai.googleblog.com/2017/11/automl-for-large-scale-image.html}}. Automated deep learning algorithms, such as AutoML, have been successfully applied for different use cases, such as medical image classification \citep{Faes2019, Korot2021}, detecting invasive carcinoma \citep{Zeng2020} or predicting COVID-19 from patients' CT scans \citep{Arellano2021}.

We found that training the AutoML model with only asteroid trails produces many false positives, as badly subtracted cosmic rays or arcs from gravitational lenses in clusters can have a similar appearance. So that the AutoML Vision model distinguishes between them, we trained the model with four labels: satellite, asteroid, gravitational lens arc, and cosmic ray (all of which produce trail-like features). Therefore, we can detect all four types of objects separately in the cutouts. Besides a classification and score, the AutoML object detection algorithm returns bounding boxes for each identified object \citep{Xin2019}.

From the visual classification of citizen scientists, we used a total of 1\,488 asteroid labels and 1\,343 cosmic ray, 698 gravitational lens arcs, and 1\,673 satellite labels (based on the tags on Talk). We split the sample into 70\% training set, 15\% validation, and 15\% test set, using a random selection. The validation set is used by AutoML to fine-tune the preprocessing, architecture and hyperparameter optimisation and to validate the results. We use the test set to assess the performance of the model we trained. Our model achieves a precision of 78.3\% and a recall (completeness) of 61.1\% on all labels, with a score threshold of 0.5 and an intersection-over-union (IoU) of 0.5. For the particular case of asteroid trails, our AutoML model achieves 73.6\% precision, 58.2\% recall (or completeness), and 65.0\% F1-score. The metrics from our model are shown in Fig. \ref{automl_metrics}.

\subsection{AutoML classifications}
\label{AutoML_classifications}

Our aim in using an automated deep learning classifier is to extend the detection of asteroid trails when new data reach the eHST archive. Our AutoML classification of 149\,292 cutouts, including the dataset classified by the volunteers and described in Sect. \ref{volunteerclassifications} and new images available in the archive as of 14 March 2021 (thus extending our analysis with data for almost one year), returns 2\,041 asteroid trails.
See examples of two such asteroid trails identified by AutoML in Fig.~\ref{examples_from_autom}. We cross-matched this set with the volunteer classifications from Hubble Asteroid Hunter. We find 1\,044 common objects, using a positional matching tolerance of 4 arcsec for WCF3/UVIS and 5 arcsec for ACS/WFC. These values are conservative, taking into account the average object detection bounding-box size yielded by the AutoML algorithm ($\approx$ 10 arcsec). There are 997 new asteroid trail candidates identified by AutoML. This brings the total number of asteroid trails detected by citizen scientists and AutoML to 2\,487.  

Asteroid trails appear in two different ways in HST ACS/WFC and WFC3/UVIS images: (1) bright trails have their central part removed by the DrizzlePac cosmic ray rejection algorithm, but with the edges of the trails left visible; (2) in the case of faint asteroids, the trails are below the threshold for cosmic ray rejection and are not affected. To calculate the threshold, the cosmic ray rejection algorithm compares the composite median image to individual exposures, taking expected alignment errors and noise statistics into account \citep{Gonzaga2012}.

Asteroid trails have a different shape compared to cosmic rays. While cosmic rays are often removed completely by the rejection algorithm, the asteroids trails, having shapes convolved with the point spread function of the telescope, are never completely removed (the edges are still visible in the composite images). Our image detection algorithm and the citizen scientists were trained using both types of asteroid trails: erased (but with edges visible) and not erased by the cosmic ray rejection algorithm. The completeness of our study is therefore dominated by the performance of our detection algorithm, which was also trained to distinguish cosmic rays that were not completely erased (see Sect. \ref{automl}). The DrizzlePac cosmic ray rejection algorithm applied to asteroids trails has a negligible contribution to the completeness of our sample.

\subsection{Trail processing}
\label{trail_processing}

We detected 2\,487 trails. All of them were visually inspected by three of the authors (SK, PGM, and MP) using the project on the Zooniverse platform, and we identified 1\,790 valid trails, 410 cosmic rays, and 287 false identifications (no trail or trails for which the identification was ambiguous, i.e. not validated by all three of us after the visual inspection). We removed 65 trails that were associated with \textit{HST} observations targeting SSOs. Of the 1\,725 remaining trails, 24  were identifications of two trails very close to each other or that intersect in the composite images. Since their trail processing and photometry requires a special treatment, we left them for further assessment and excluded them from this dataset. Our final dataset consists of 1\,701 trails, found in 1\,316 \textit{HST} composite images.

The next step of data processing was to retrieve the astrometric positions --  right ascension (RA) and declination (Dec) -- and the total flux. Once this information was computed, we could identify the trails corresponding to known SSOs. To perform these two tasks, we designed two software pipelines: (1) the \emph{Trail Extractor}, used to extract the trail from the cropped images and to compute the astrometric and photometric properties; and (2) the \emph{Trail Matching} pipeline, used to identify the corresponding SSOs.

The cropped images with asteroid trails used for \textit{Trail Extractor} were generated by going back to original raw \textit{.fits} images and re-processing them with DrizzlePac, overriding the cosmic ray rejection step, which may delete a considerable number of pixels in the trail (see Sect. \ref{AutoML_classifications}).

\begin{figure}
   \centering
   \includegraphics[width=1.2cm]{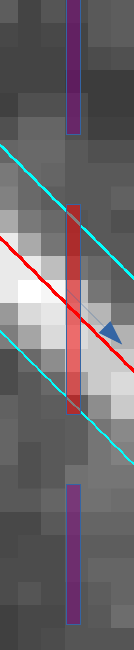}
   \caption{Schematic of the trail extraction. The red line shows the middle longitudinal section of the trail. The red rectangle shows the aperture (bordered by the two cyan lines) for computing the flux. The violet regions are used for estimating the background value.}
    \label{TrailIdent}%
\end{figure}

The {\bf Trail Extractor} pipeline is used to retrieve the trail from the \emph{.fits} files and to obtain the calibrated astrometric and photometric data. First, the (x,y) position and the flux, provided in terms of analogue to digital units (ADU), are determined for each point along the trail path. The algorithm is implemented using GNU Octave software \citep{gnuoctave}. Its schematic is shown in Fig.~\ref{TrailIdent}.  The first task is to determine the trail and its centre. This is done by finding the maxima along each column from the image cutout (in some cases the image is transposed in order to have its longest dimension on the x-axis, i.e. the number of columns larger than the number of rows). An outlier removal procedure is used in order to keep only the pixels belonging to the trail. A curve that marks the centre of the trails is traced along the (x,y) positions of the pixel maxima using the \emph{splinefit} function from GNU Octave. 

\begin{figure}
   \centering
   \includegraphics[width=7cm]{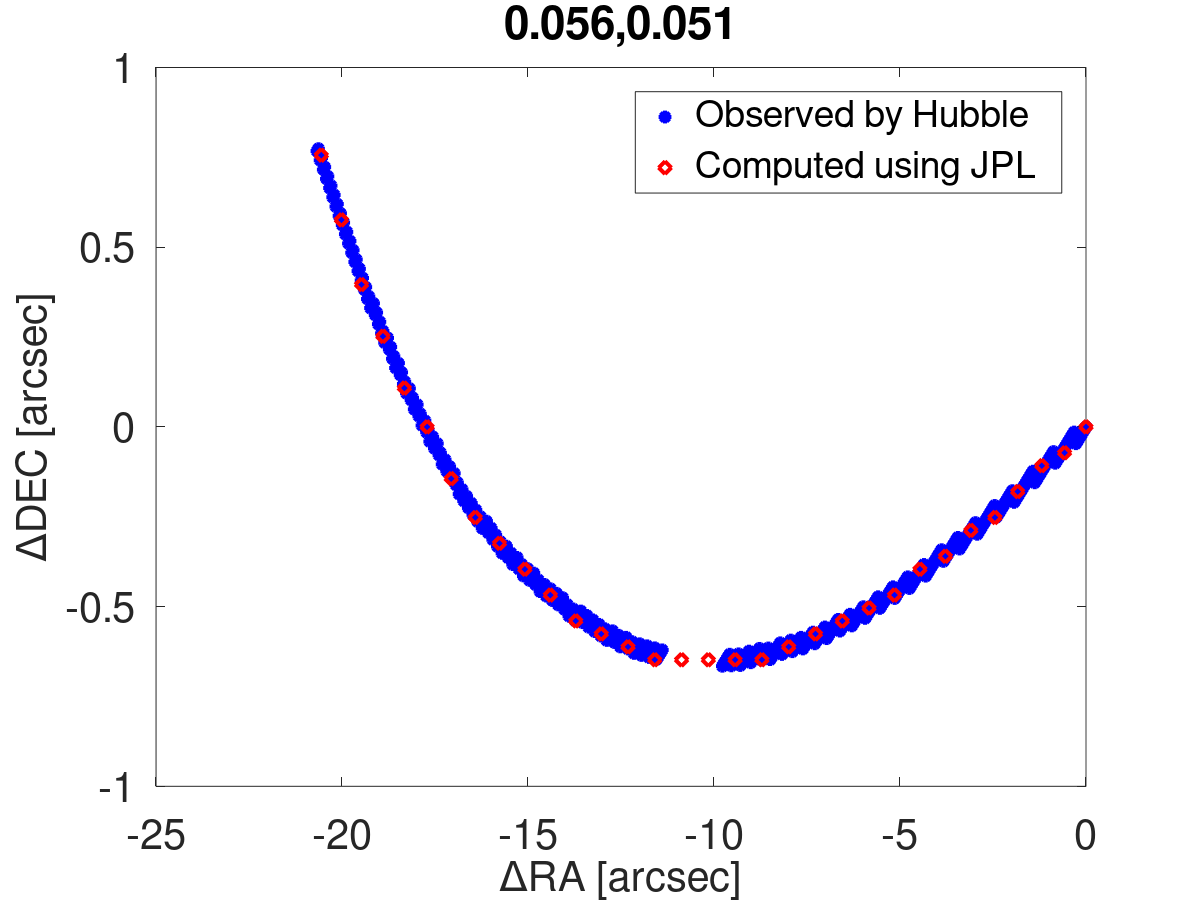}
   \caption{Example of matching between the trail observed by \textit{HST} and the ephemerides data computed with a time step of one minute using the JPL Horizon web service. In this case, the Main-Belt asteroid (167339) 2003 UN308 was identified. The title shows the median and the average of the observed minus predicted position of the asteroid (the unit is arcseconds).}
    \label{TrailMatch}%
\end{figure}

An aperture of nine pixels wide is used to determine the flux (this is done by summing the pixel values) on each column by following the trail centres. The background is estimated as a median value of the adjacent pixels (as shown in Fig.~\ref{TrailIdent}) and then subtracted. The (x,y coordinates) corresponding to trail centres are converted to (RA, Dec) using the world coordinate system (WCS) parameters provided in the \emph{.fits} file. The reported astrometric precision of the WCS solution in the {\it HST} images used for this work is 0.3 arcsec or better, according to Space Telescope Science Institute (STScI)\footnote{\url{https://outerspace.stsci.edu/pages/viewpage.action?spaceKey=HAdP&title=Improvements+in+HST+Astrometry}}. The total flux of the trail (\emph{F}[ADU]) expressed in electrons per second is converted to magnitude. Equation~\ref{eq:mag} describes this transformation\footnote{\url{https://www.stsci.edu/hst/instrumentation/acs/data-analysis/zeropoints}}: 

\begin{equation}
\begin{aligned}
m_{AB} = & -2.5\log_{10}(F) -2.5\log_{10}(PHOTFLAM) \\
         & - 2.5\log_{10}(PHOTPLAM) - 2.408,
\end{aligned}
\label{eq:mag}
\end{equation}

\noindent{where \emph{PHOTFLAM} is the inverse sensitivity and represents the scaling factor necessary to transform an instrumental flux in units of electrons per second to a physical flux density, and \emph{PHOTPLAM} is the pivot wavelength provided in the header and used to derive the instrumental zero-point magnitudes. This formula is applied correctly if the object was in the FoV during the entire exposure. Otherwise, we had to apply a scaling factor proportional with the time interval during which the object was in the exposure.} 

The \emph{TrailExtractor} method works well for trails with high signal-to-noise ratio of individual points (e.g. larger than 5). It can fail for trails that have a low signal-to-noise ratio (smaller than 3). For those where it failed (in about 33\% of the cases), we manually provided some additional positions along the faint line in order to help the algorithm identify it. This task was done using the Zooniverse interface during the visual inspection of all trails by the authors.  

\subsubsection{Timing accuracy}
\label{time_accuracy}

To determine the sky position of an SSO, we need to know the precise moment of its observation. Thus, the error for determining the beginning and the end (the margins) of a trail translates into a time uncertainty for the reported trail. These possible errors are of three types. The first is the error introduced by the detector noise. This error is negligible for trails with high signal-to-noise ratio, but it can be misleading for trails at the detection limit. The second is the error introduced due to background sources or cosmic rays at the edges of the trail.
The third is the error introduced if the object was not in the FoV for the entire exposure (i.e. the object entered in or went out of the FoV during the image exposure). This is the most significant error as we do not know for how long the object was imaged.

We identified trails in the composite exposures. In order to quantify the total time (\emph{trailexp}) when the SSO was in the FoV we examined the individual images for the appearance of corresponding trail. The fact that \emph{trailexp} time could be smaller than the total exposure time of the stacked image (\emph{imageexp}) translates into a magnitude correction for Eq.~\ref{eq:mag}. This is equal to

\begin{equation}
\begin{aligned}
    \Delta m = 2.5 \times \log_{10}\left(\frac{trailexp}{imageexp}\right).
\end{aligned}
\label{eq:m}
\end{equation}

\subsubsection{Identifying the Solar System bodies (TrailMatching)}
\label{TrailMatching}

In order to identify the known SSOs that generated these trails, we need the accurate time of the observation and the accurate position of the object.  As a first approximation, we took the average position of each trail and the moment corresponding to the middle of the exposure time of the composite image. This information was used as input for SkyBoT\footnote{\url{https://ssp.imcce.fr/webservices/skybot/}} \citep{SkyBoT2006,Berthier2016}. Using this tool, we performed a cone search, with a radius of 900 arcsec (this larger field was used in order to account for all various uncertainties), for all asteroids potentially associated with this trail. 

Because of the wide-cone used for the search, a number of 22\,748 known SSOs were found as possible candidates. For all these candidates we retrieved the predicted ephemerides using the online \emph{Jet Propulsion Laboratory (JPL) Horizons} service\footnote{\url{https://ssd.jpl.nasa.gov/horizons/}}, considering the orbital position of \textit{HST}. The ephemerides were obtained using a step of 1 min for the entire interval corresponding to the composite exposure. Then we searched for the best match between the predicted coordinates for each SSO and the corresponding trail coordinates observed by \textit{HST} by using the approximate distance formula, $d_{O-C}$ (Eq.~\ref{eq:angdist}). We assumed a 3\% error in the exposure time (this value was selected as a rough estimation to account for the precision in identifying the start of the trail) and searched for SSO matches within this error. Since we do not know the direction of the asteroid's motion \emph{a priori}, we considered both directions in the matching:

\begin{equation}
\begin{aligned}
& d_{O-C} = \sqrt{[\cos(Dec_{pred})\cdot (\Delta RA)]^2 + (\Delta Dec)^2} \\
& \Delta RA = RA_{pred} - RA_{obs} \\
& \Delta Dec = Dec_{pred} - Dec_{obs} \\
\end{aligned}
\label{eq:angdist}
.\end{equation}
Thus, by considering a threshold $d_{O-C} \leq 3\arcsec$, we found 670 matches (trails associated with known SSOs). These correspond to 454 unique known SSOs. An example of a successfully matched trail is shown in Fig.~\ref{TrailMatch}.

The reason for using a 3$\arcsec$ tolerance to identify the known objects is not solely due to the WCS astrometric precision (0.3$\arcsec$; see Sect. \ref{trail_processing}).\ It is also related to the accuracy of identifying the beginning and the end of the trails (in particular for the low signal-to-noise ones), as described in Sect. \ref{time_accuracy}.

\section{Results}
\label{resultssection}

Using the two methods to detect asteroids in the \textit{HST} images, citizen science and deep learning, we identified a final sample of 1\,701 trails corresponding to SSOs in 1\,316 composite \textit{HST} images. Of these, 670 trails correspond to known SSOs. We refer to the sample of 670 trails as `identified' asteroids and remaining 1\,031 trails as `unidentified'. We present the results of our search for SSO trails in the \textit{HST} archives, including identified trails (with the associated SSO) and the unidentified ones in Table \ref{FullTable}.

\begin{figure}
   \centering
   \includegraphics[width=\columnwidth]{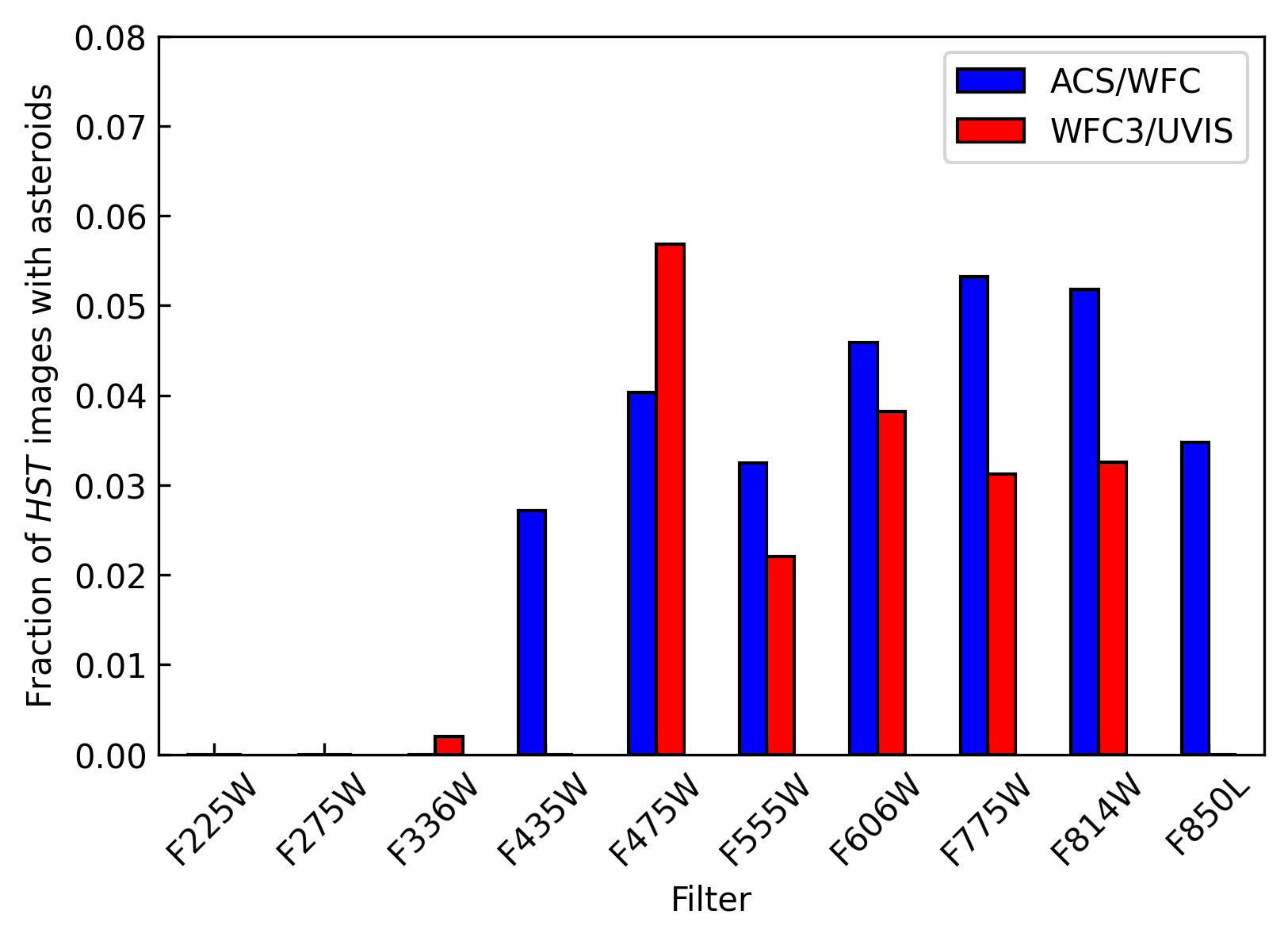}
   \caption{Fraction of \textit{HST} composite images with detected asteroids, split by filter (for the eight most common filters for each instrument and for the two instruments, ACS/WFC and WFC3/UVIS.}
    \label{asteroid_filter}%
\end{figure}

We find, on average, asteroid trails in 1\% of the cutouts or in 3.5\% of the \textit{HST} composite images available in the eHST archive. We plot the fraction of \textit{HST} composite images with detected asteroids as a function of the \textit{HST} filter used for observations in Fig. \ref{asteroid_filter}. When split by filter, we find that the V (F475W, F555W), R (F606W) and I bands (F775W, F814W) contain the highest fraction of objects, 5\%. The UV filters (<400nm) contain the lowest fraction of asteroids, <1\%, despite imaging similar regions of the sky. This implies that it is difficult to identify asteroids in the UV, where SSOs are less bright and the magnitude limit of the observations is lower. It is not unexpected that fewer SSOs are observed in the blue and UV filters as asteroids reflect the incident sunlight and the solar irradiation intensity is largest in the visible (peaking in the V bands).

\begin{figure}[h]
     \centering
     \begin{subfigure}[a]{0.5\textwidth}
       \centering
       \includegraphics[width=\columnwidth]{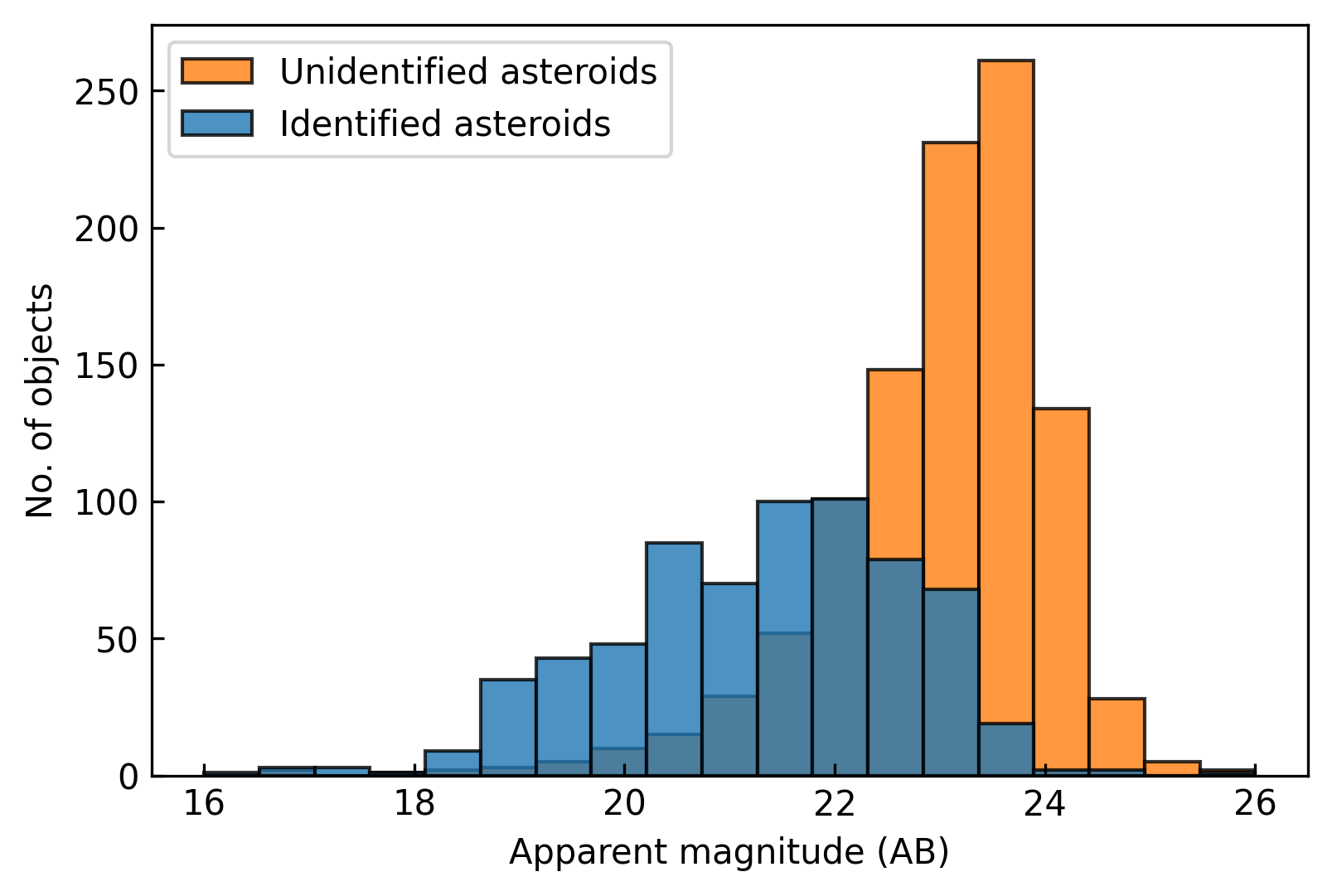}
     \end{subfigure}
     \hfill
     \begin{subfigure}[b]{0.5\textwidth}
       \centering
       \includegraphics[width=\columnwidth]{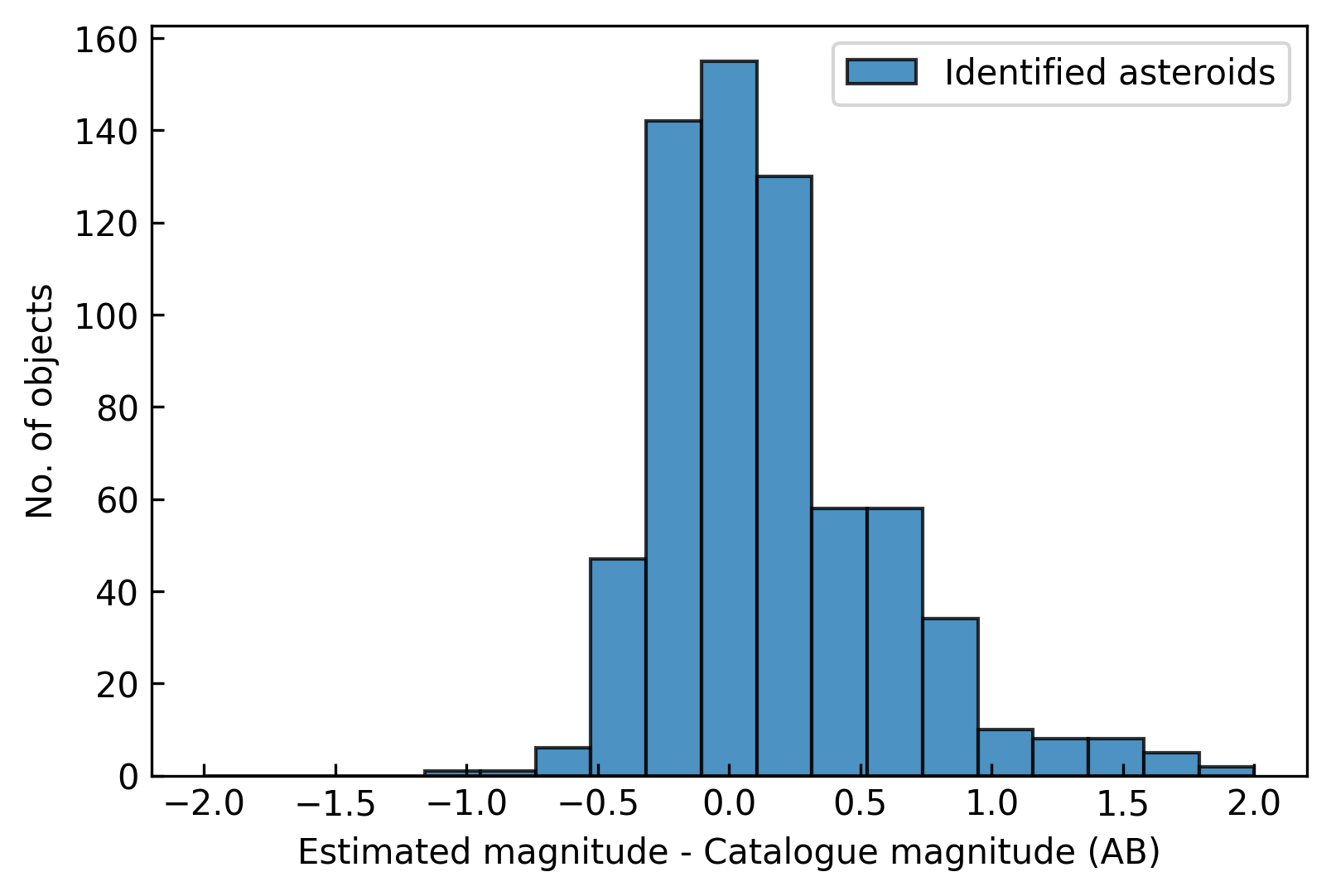}
     \end{subfigure}
     \caption{Distribution of apparent magnitudes for the SSOs identified in the \textit{HST} images. (a) The measured magnitudes for the identified objects (blue bars) and for the objects for which we did not find any associations with known SSOs (orange bars). (b) Difference between the measured magnitude in the filter of the \textit{HST} observation and the V-band magnitude in the MPC database for the identified asteroids.}
     \label{appmag}
\end{figure}

We plot the distribution of the measured apparent magnitudes of the objects in Fig. \ref{appmag}a. The measured magnitudes for the identified asteroids match the published values in the MPC database well, as shown in Fig. \ref{appmag}b. The scatter is due to using different filters to identify asteroids in \textit{HST}, whereas the MPC magnitudes are reported for the V band (centred around 0.55\,$\mu$m). The V-band magnitude of the asteroids observed in the \textit{HST} images cannot be computed with certainty due to the variable spectral energy distribution of asteroids based on their surface composition. As such, we do not attempt to convert the observed magnitudes into the catalogue ones. The distribution of the differences in magnitudes is skewed towards positive differences as many asteroids exhibit a strong decline in spectral reflectance towards 0.9\,$\mu$m, close to the observation wavelengths of the \textit{HST} exposures \citep{Demeo2009}.

It is immediately clear in Fig. \ref{appmag}a that the unidentified asteroids have fainter magnitudes compared to the those matched with known SSOs. The identified asteroids have a broad distribution of magnitudes, with a median of 21.5 mag and $\sigma$ of 1.4 mag, whereas, the median magnitude for the unidentified objects is 1.6 mag fainter, 23.1 mag and $\sigma$ of 1.1 mag. While there is an overlap between the two distributions, there are significant differences: only 20\% of objects fainter than 22 mag have been identified with known SSOs. On the other hand, 17\% of objects with magnitudes brighter than 21 mag have not been identified. This demonstrates the potential of \textit{HST} to image and detect unknown, faint asteroids. 

\begin{figure*}
   \centering
   \includegraphics[width=\textwidth]{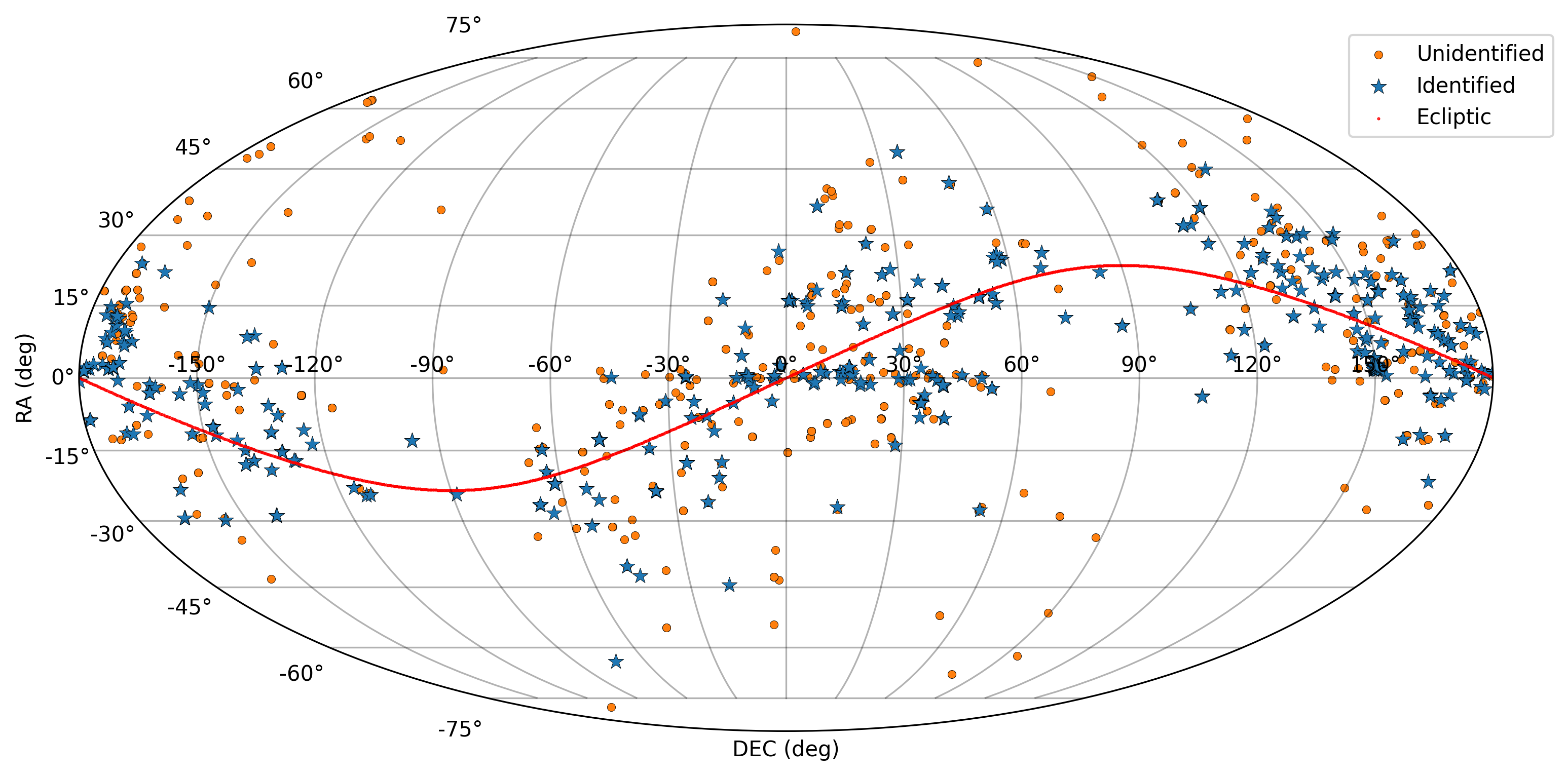}
   \caption{Distribution on the sky of the SSOs identified in the \textit{HST} images in Mollweide projection. The blue stars show the identified, known asteroids. The orange circles show the location of objects for which we did not find any associations with SSOs. The ecliptic is shown with red. The two gaps in this plot correspond to the Galactic plane, which was not observed by \textit{HST}.}
    \label{ravsdec}%
\end{figure*}

\begin{figure}[h]
     \centering
     \begin{subfigure}[b]{0.5\textwidth}
       \centering
       \includegraphics[width=\columnwidth]{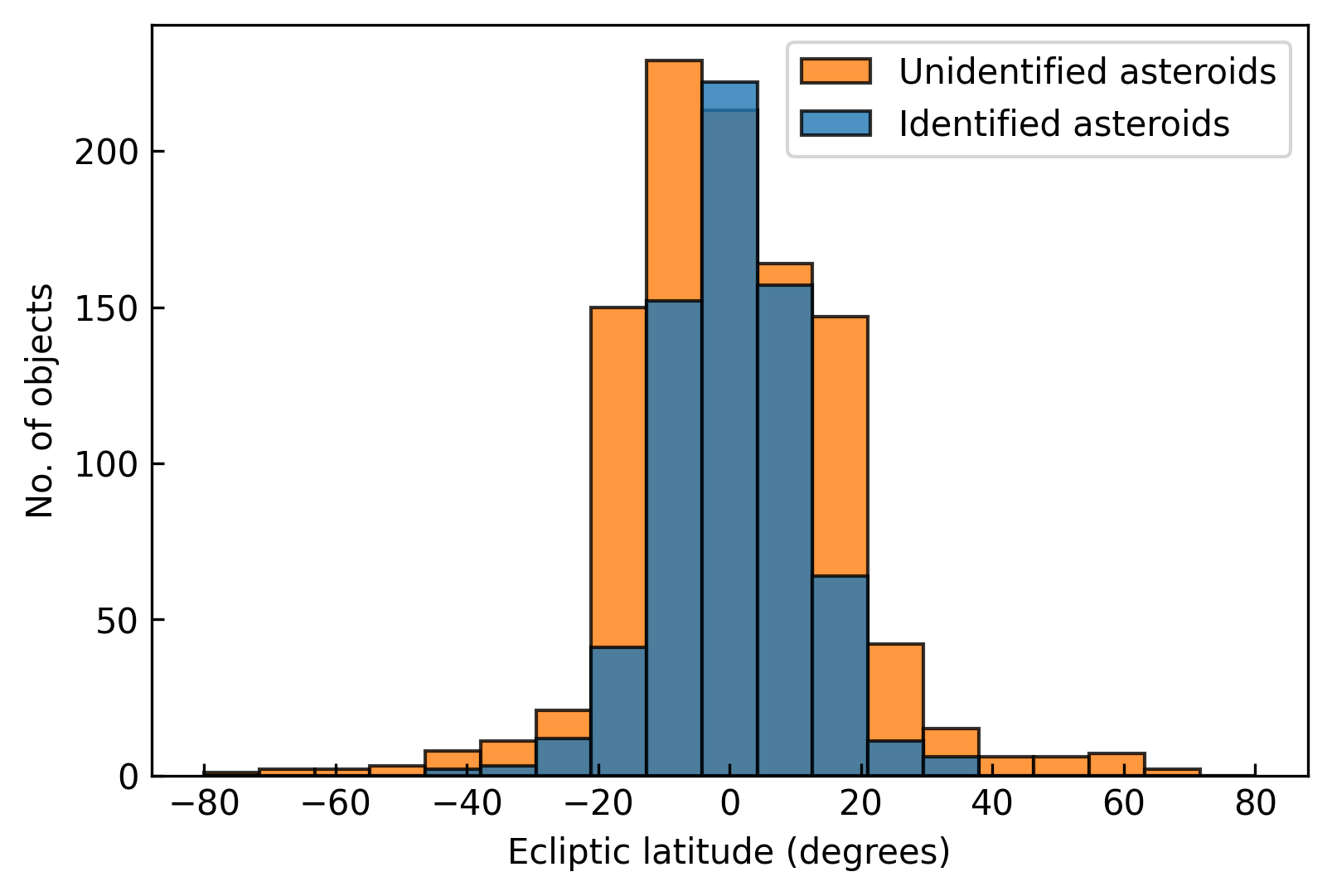}
     \end{subfigure}
     \hfill
     \begin{subfigure}[b]{0.5\textwidth}
       \centering
       \includegraphics[width=\columnwidth]{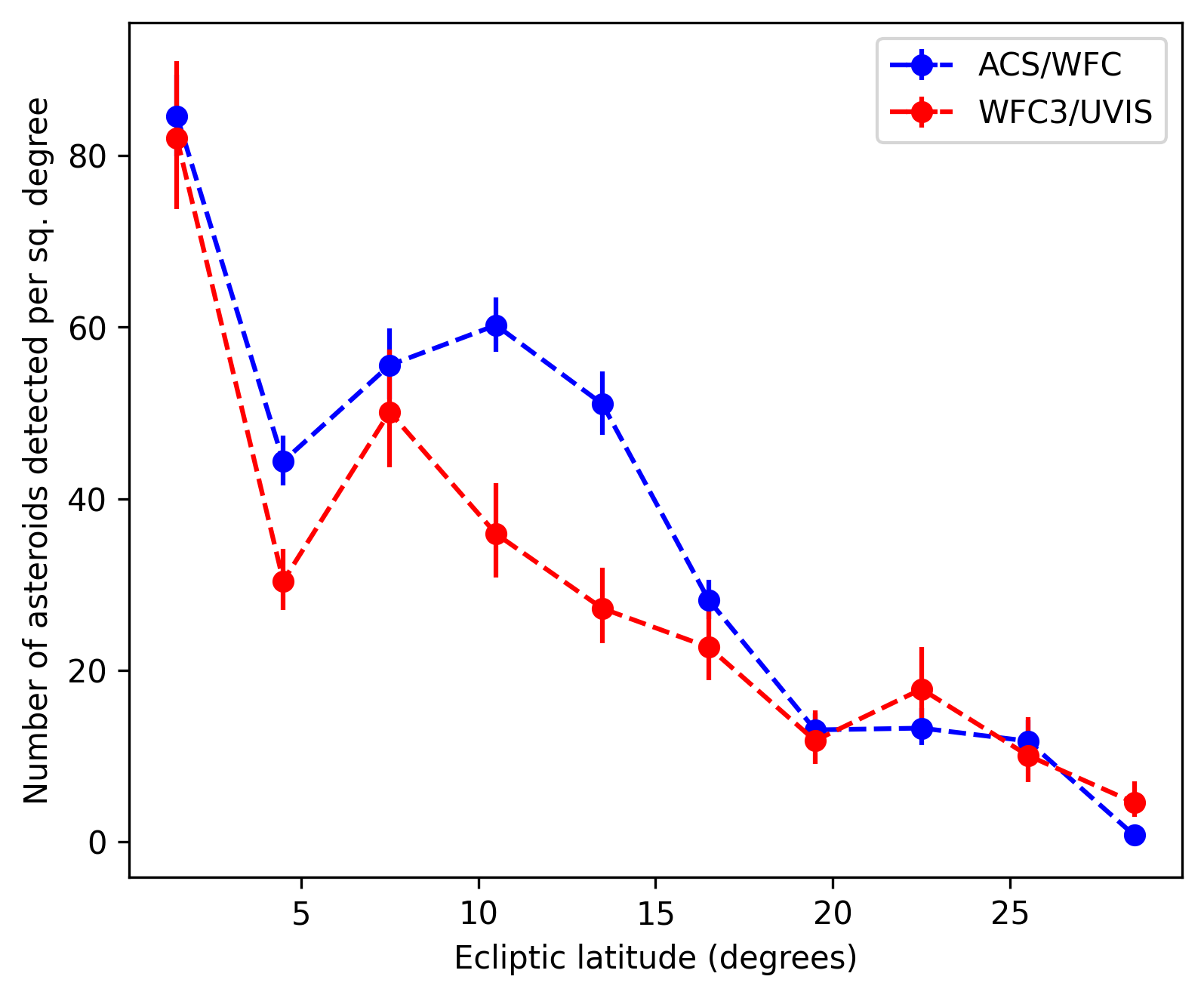}
     \end{subfigure}
     \caption{Distribution of asteroids identified in \textit{HST} images by ecliptic latitude. (a) Number of identified (blue) and unidentified (orange) objects by ecliptic latitude. (b) Number density of detected asteroids with magnitudes < 24.5 mag, as a function of ecliptic latitude, for the two \textit{HST} instruments.  The errors are the 68\% (1$\sigma$) confidence limits from the \citet{Wilson1927} binomial confidence interval.}
     \label{ecliptic_latitude_distribution}
\end{figure}

Next, we investigated the distribution on the sky of detected SSOs. In Fig. \ref{ravsdec} we plot the RA and Dec of the asteroids, split into known and unknown SSOs. Most asteroids depict small ecliptic latitudes, as expected for the products of the collisional processes during the formation of the Solar System \citep{Morbidelli2015}. In Fig. \ref{ecliptic_latitude_distribution}a we plot the distribution of the ecliptic latitudes ($b$) of the observed asteroids. Unidentified asteroids show a wider distribution of ecliptic latitudes ($\sigma=17^{\circ}$ vs. $\sigma=10^{\circ}$) similar to the report by \citet{Mahlke2018_SSO} and \citet{Carry2021} on SSO discoveries with the Kilo-Degree Survey (KiDS) and Gaia, respectively. There are only seven known asteroids at $|b|>30^{\circ}$ and 64 unidentified asteroids at $|b|>30^{\circ}$. 6\% of all the unidentified ones are at high ecliptic latitudes ($|b|>30^{\circ}$), suggesting highly inclined orbits. This shows the potential of \textit{HST} in observing small asteroids with high inclinations. In Fig. \ref{ecliptic_latitude_distribution}b we plot the sky density of detected asteroids as a function of ecliptic latitude. We detect 55 asteroids up to a magnitude of 24.5 mag per square degree up to $b=15^{\circ}$, on average, in the \textit{HST} archives (58 in ACS/WFC and 41 in WFC3/UVIS). As expected, the density decreases from $\sim$80 asteroids per square degree at $b=0^{\circ}$ to $\sim$1 asteroid per square degree at $b>30^{\circ}$. This value is smaller compared to the result of \cite{Heinze2019} who found a density of $455\pm13$ asteroids by using the Dark Energy Camera (DECam) mounted on the 4 m Blanco Telescope. The difference is explained by the observing strategy. They observed a single field at opposition from the Sun (where we expect the observed asteroid density to be maximum), using broadband filters (mostly the V and R). Our results are obtained using \textit{HST} observations taken at various phase angles, mostly with narrow-band filters covering various wavelength ranges, and therefore without the optimal conditions to image asteroids.

Finally, we investigate the sky motion (differential rates) of the asteroids we found in the \textit{HST} images. We calculate the sky motion for each object with the measured (curved) trail lengths and the time interval between the start of the first exposure and the end of the last exposure used to obtain the stacked image. Because \textit{HST} moves along its orbit during the exposure, the parallax effect must be taken into account. For a Main-Belt asteroid, this is of the order of $\sim$ 10 arcsec, which is comparable with the trail lengths we are detecting. The parallax effect, and the properties that can be derived from it, will be analysed in a future article.

\begin{figure}
   \centering
   \includegraphics[width=\columnwidth]{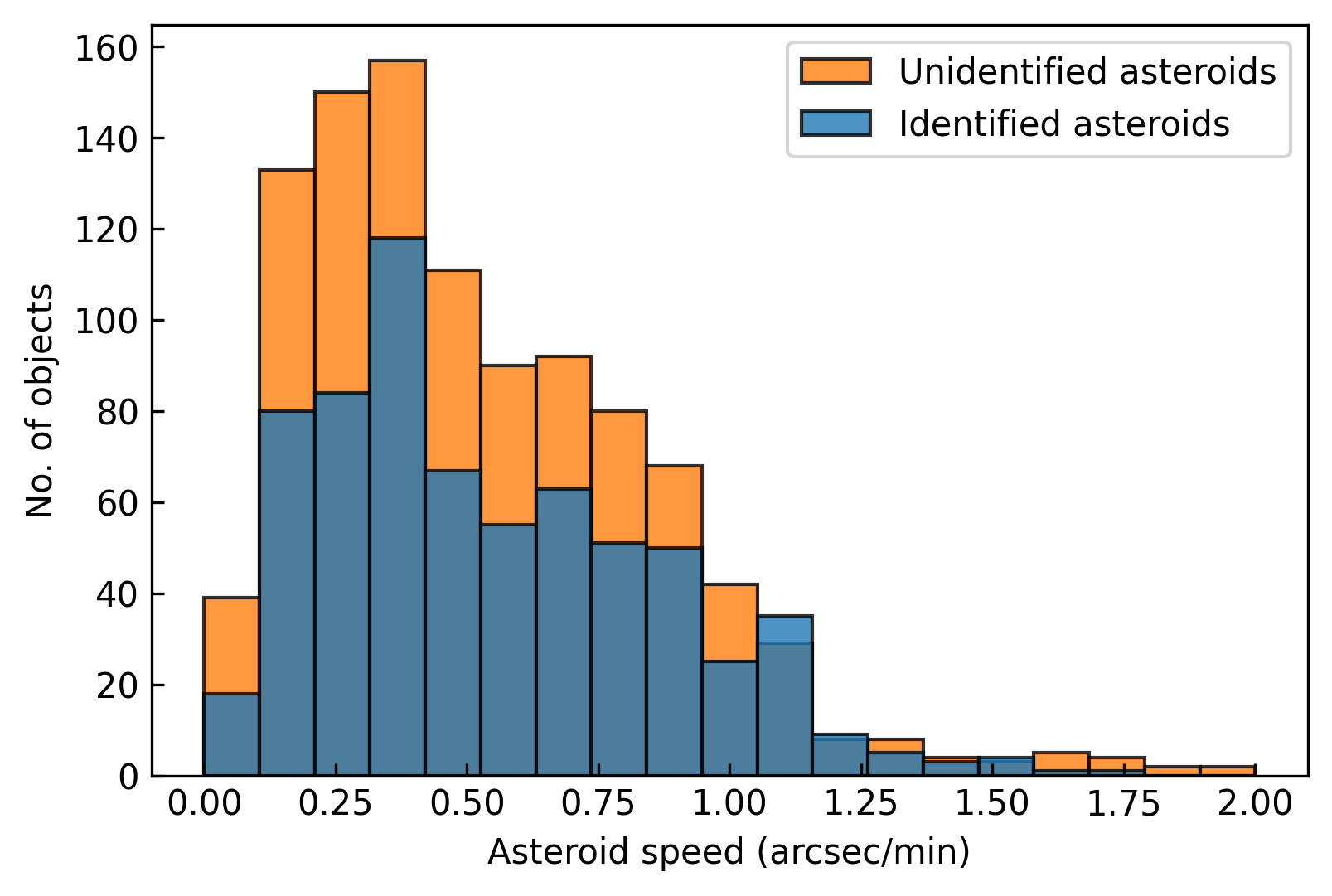}
   \caption{Distribution of the sky motion (in arcsec/min) for the known and unknown SSOs in the \textit{HST} images.}
    \label{speed}%
\end{figure}

We plot the distribution of sky motions in Fig. \ref{speed}. This distribution is confined to between 0.01 and 2 arcsec/min, with only eight objects having an average differential rate larger than 3 arcsec/min, all of which are unknown. Their sky motion is typical for near-Earth asteroids (NEAs).   

Near-Earth asteroids are fast moving objects that could cross the entire FoV and, in principle, be misidentified with satellite trails. This may occur for apparent motions larger than 16-20 arcsec/min (estimated by taking into account the size of the FoV and the median exposure time of $\sim$10 min for a single image; although the asteroid may be imaged in a corner of the detector and correspondingly the threshold would be lower).  These large apparent motions for NEAs occur when they are at several lunar distances from Earth, or less, which is known to be rare. Additionally, even high apparent motion NEAs are expected to be affected by parallax and produce curved trails, whereas the detected satellite trails are straight (they are crossing the FoV in a few seconds). In most cases, satellite trails (studied in a dedicated future publication) are easily distinguishable from asteroids because they are bright, cross the entire FoV and do not show the parallax effect. 

The lower limit (0.01 arcsec/min) of this histogram is given by our trail detection method. Although we found 20 trails that correspond to a differential rate less than 0.1 arcsec/min, which may correspond to distant, Centaur-type asteroids or trans-Neptunian objects (TNOs), we did not identify any of these trails with known distant objects. We were able to identify some of the long-period comets and TNOs that were targeted on purpose by \textit{HST} (i.e. with optimal observing conditions). As these were targeted observations, they were not included in our list.  The main reason for the absence of short trails associated with TNOs is the detection method used, which is not optimised for very short trails (less than $\sim$20\ pixels). Thus, any object that had an apparent trail shorter than $\sim 1$ arcsec will likely not be detected (it will be considered as a background source). 

\begin{figure}
   \centering
   \includegraphics[width=\columnwidth]{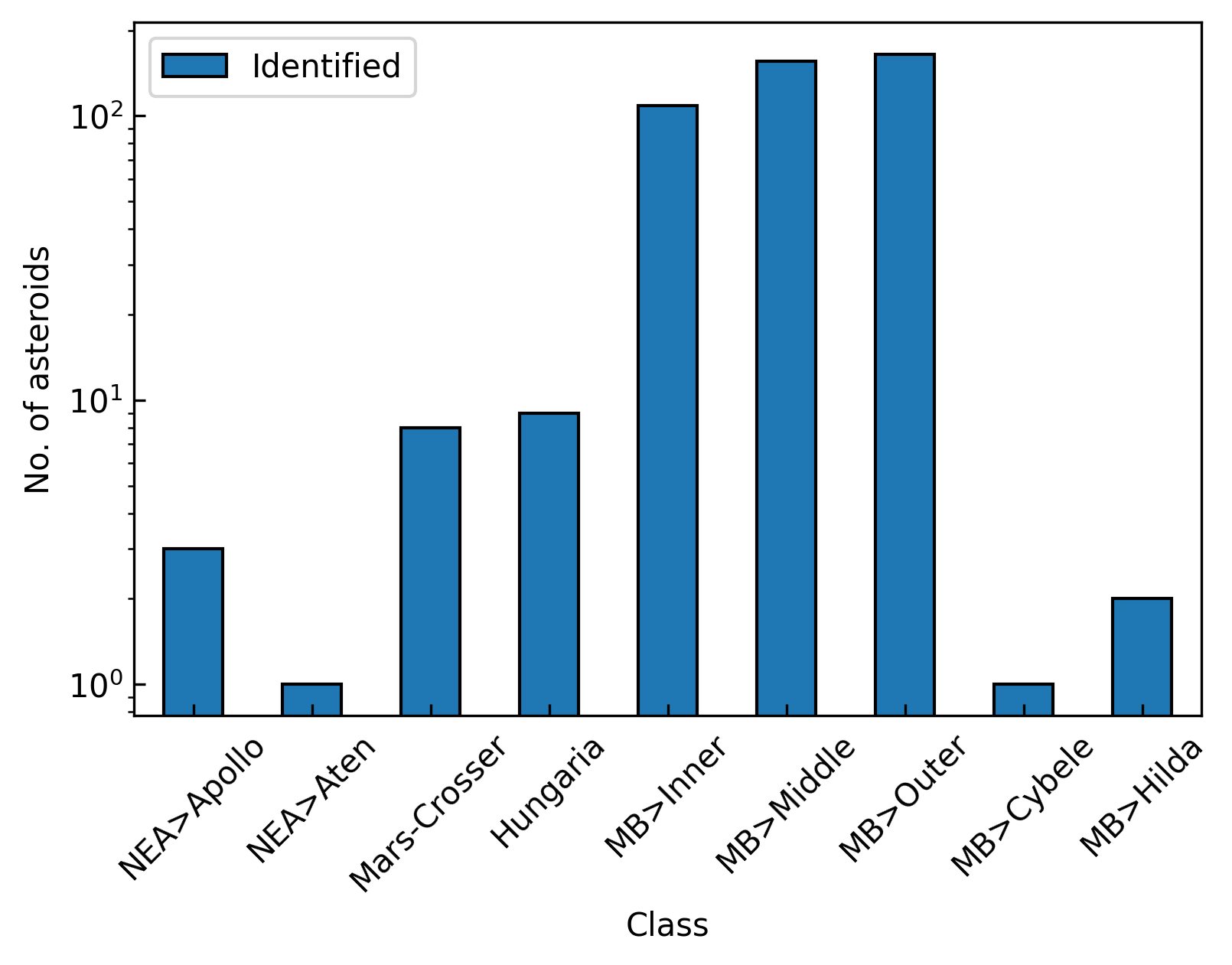}
   \caption{Distribution of asteroid orbital classes for the 454 objects identified with known SSOs. The asteroid classes are sorted in ascending order based on the semi-major axes of their orbits.}
    \label{classes}%
\end{figure}

\subsection{Identified objects}

We identified 670 trails with known SSOs, as described in Sect. \ref{TrailMatching}. They correspond to 454 individual SSOs, 147 asteroids appearing in more than one image (some consecutive exposures contain the trails of the same asteroid). Thirty-three of them appear in at least three different images. We plot the distribution of the identified asteroid orbital classes in Fig. \ref{classes}. Out of these, 433 (95\%) are Main-Belt asteroids (inner, middle, outer, Cybele, or Hilda) and only 21 (5\%) are of other types, including nine Hungarias, eight Mars-crossers, and four NEAs. The deep view of the \textit{HST} instruments is once more apparent: the most observed orbital population is the outer Main Belt, where asteroids are more numerous, farther away, and in general darker than the objects in the inner Main Belt \citep{DeMeoCarry2014}.

\begin{figure}
   \centering
   \includegraphics[width=\columnwidth]{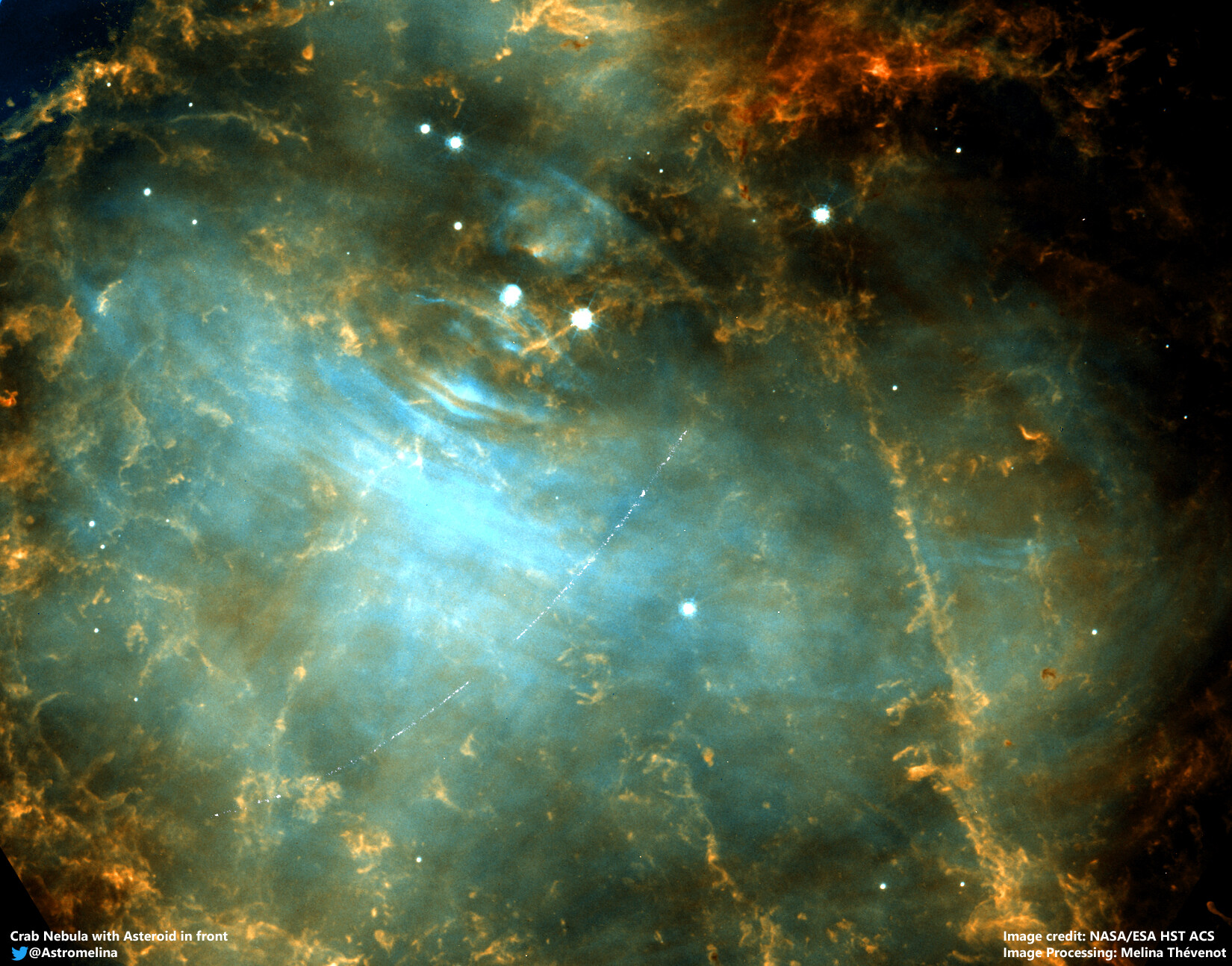}
   \caption{Asteroid (190838) 2001 SE101 passing in front of the Crab Nebula (M1) in observations with the \textit{HST} ACS/WFC F550M band (observation j9fx11010), taken on 5 December 2005. The pseudo-colour composition with F606W and F550M filters was generated by citizen scientist Melina Th\'evenot.}
    \label{crab}%
\end{figure}

An example of a known asteroid identified by citizen scientists in the project\footnote{And highlighted as an ESA image of the week - \url{https://www.esa.int/ESA_Multimedia/Images/2019/10/Foreground\_asteroid\_passing\_the\_Crab_Nebula}}, which passed in front of the Crab Nebula (M1) on 5 December 2005, is shown in Fig. \ref{crab}. It was identified as Main-Belt asteroid (190838) 2001 SE101, discovered by the ground-based LINEAR survey in 2001.

\begin{figure}
   \centering
   \includegraphics[width=\columnwidth]{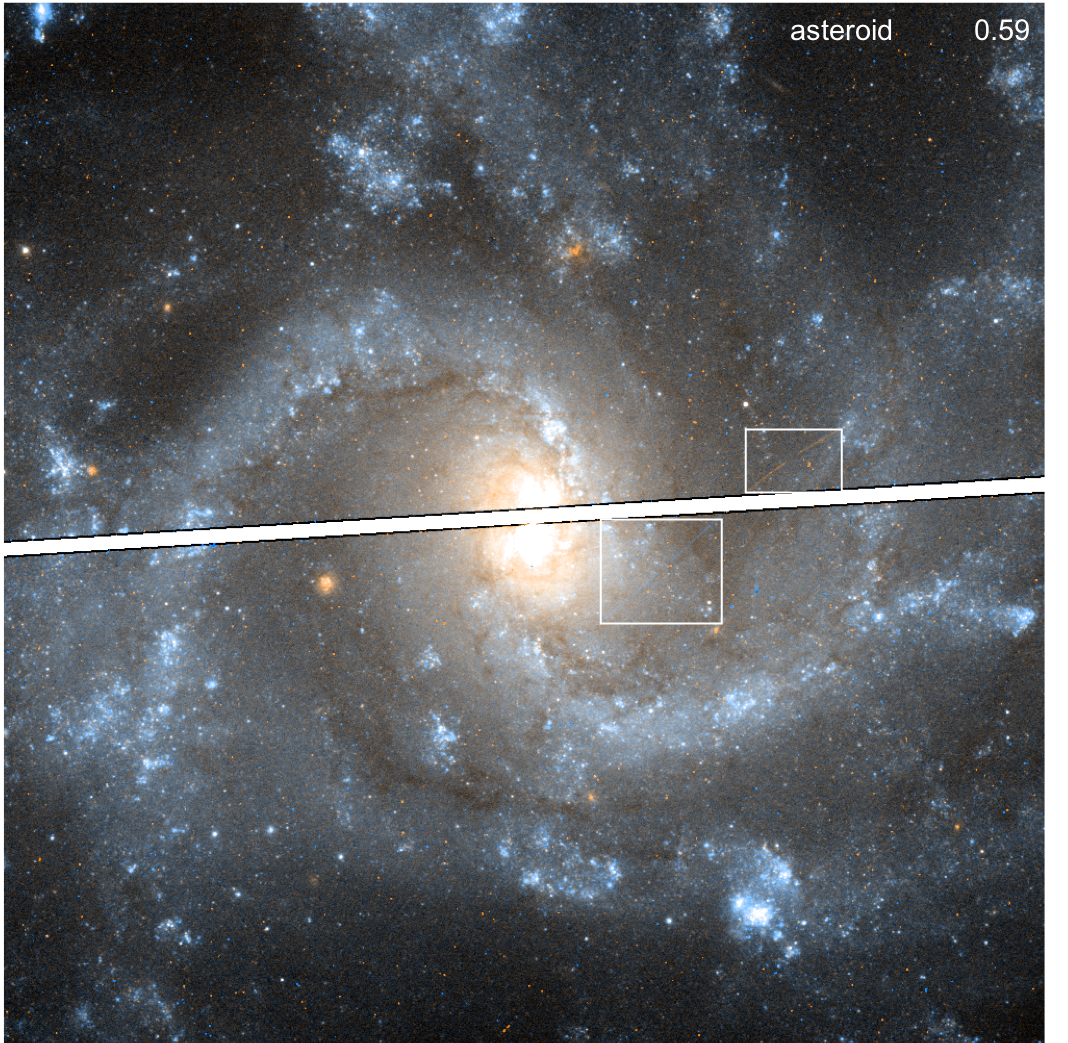}
   \caption{Asteroid (213568) 2002 LX55 moving between two exposures in front of galaxy NGC 5468 in observations with the \textit{HST} WFC3/UVIS F555W and F814W bands (observations idgg52020 and idgg52030), taken on 29 December 2017. This asteroid was detected with AutoML. The colour composition with F555W and F814W filters was generated by citizen scientist Claude Cornen.}
    \label{2002LX55}%
\end{figure}

Another example of an asteroid, identified as Main-Belt asteroid (213568) 2002 LX55, detected using AutoML, passing in front of the nearby spiral galaxy NGC 5468 on 29 December 2017 is shown in Fig. \ref{2002LX55}. As \textit{HST} was not designed to survey SSOs, this dataset contains asteroid trails passing in front of some famous extragalactic targets such as nebulae, nearby galaxies, or galaxy clusters. 

The stacked images are derived from individual exposures obtained in time intervals of 30-60 min (on average). Thus, it is possible to observe the variation in the flux along the asteroid trails and, therefore, extract the light curve of the asteroids.

\begin{figure}
   \centering
   \includegraphics[width=\columnwidth]{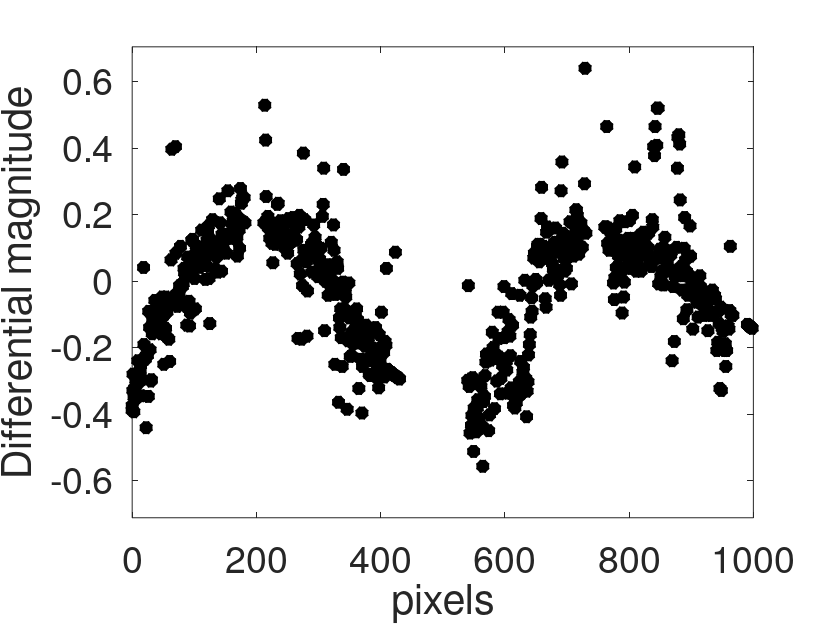}
   \includegraphics[width=\columnwidth]{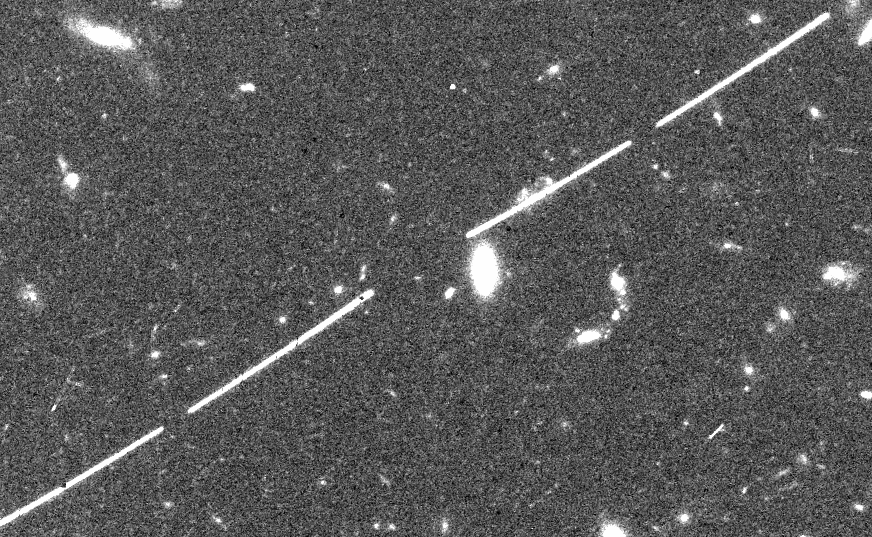}
   \caption{A possible scientific use case of the trails identified in \textit{HST} images. (a) Light curve of asteroid (114755) 2003 HP44. (b) Corresponding four trail segments.  The data correspond to a time span of 135 min. }
    \label{lightcurveexample}%
\end{figure}

In Fig. \ref{lightcurveexample}, we plot the light curve of Main-Belt asteroid (114755) 2003 HP44, varying in magnitude $\Delta m=0.6$ mag over a time-span of 135 minutes. Taking into account the median rotation period for the asteroids, which is 6.3 h (value computed by using the data available in  Asteroid Lightcurve Database\footnote{\url{https://minplanobs.org/mpinfo/php/lcdb.php}}), we can study the light curve of asteroids for $\sim15\%$ of their period. We defer the analysis of light curves for a future study.

\subsection{Unidentified objects}

\begin{figure*}
   \centering
   \includegraphics[width=\textwidth]{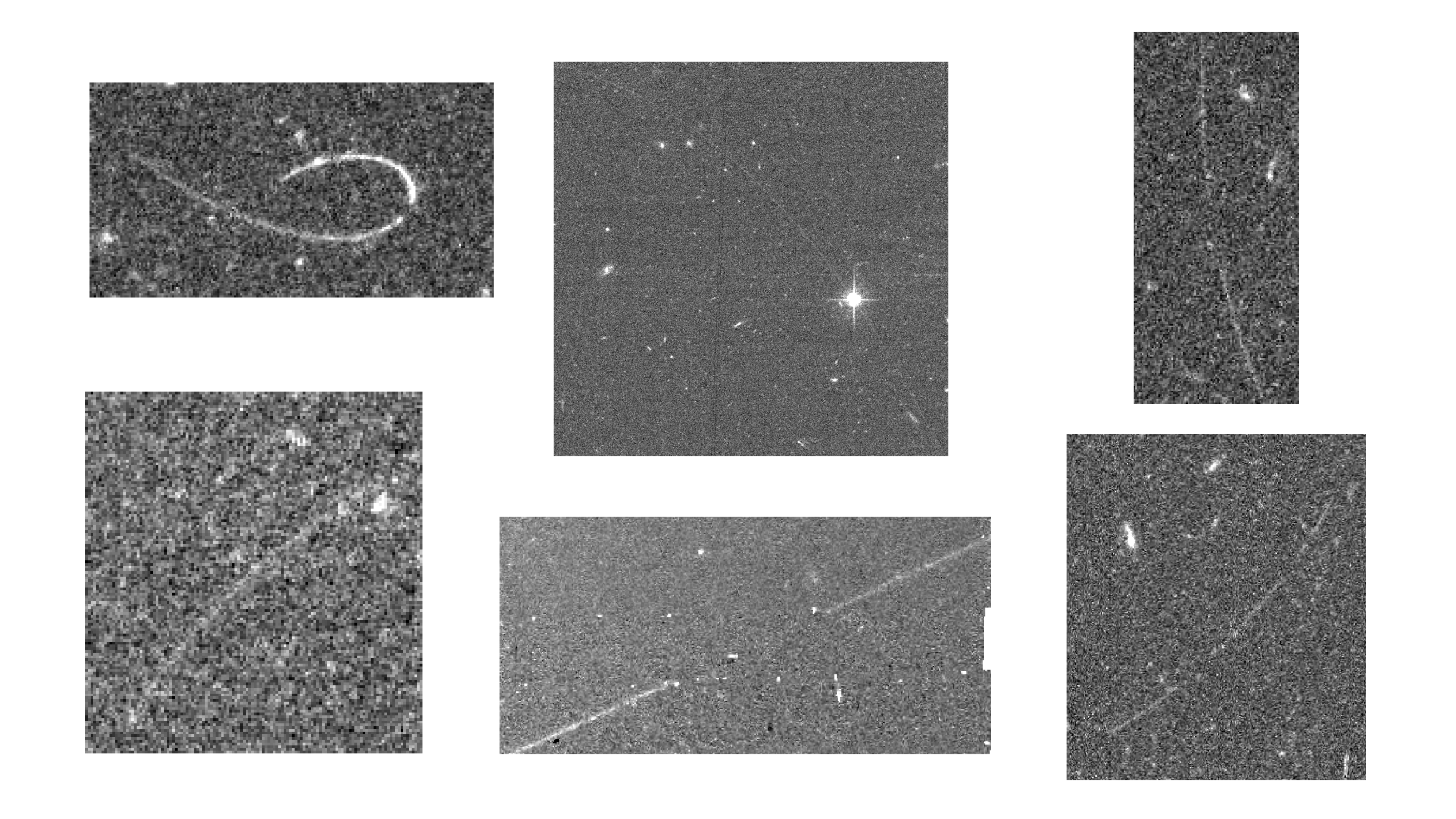}
   \caption{Examples of unidentified trails in \textit{HST} observations. The \textit{HST} observation IDs, clockwise, from the top left, are: j8pv03020, jds47w010, j9bk75010, icphg2010, jdrz23010, and jcng06010.}
    \label{unknown_asteroids}%
\end{figure*}

Unidentified objects are currently being analysed (see Sect. \ref{further_work}), and so far we can only evaluate them using their apparent magnitude, speed, and trail shape. The distributions of sky motions, shown Fig. \ref{speed}, are the same for the known and unknown objects. Therefore, we expect the unidentified objects to have a similar distribution in the orbital classes as the known ones (Fig. \ref{classes}). Nevertheless, the distributions of apparent magnitudes (in Fig. \ref{appmag}a) shows that they are fainter and, thus, smaller objects. The alternative scenario (same size objects but at larger distances) should give a different proper motion distribution. We show some interesting unidentified asteroids in the eHST archives in Fig. \ref{unknown_asteroids}, together with the associated ID number of the \textit{HST} observation. 

We discuss a few objects that present interesting features. Some trails present odd parallax shapes, for instance `S' shapes or almost a closed loop, such as the trail in j8pv03020. These special features may be helpful to constrain the orbital parameters from the parallax-induced curved trail shape (see Sect. \ref{further_work}). Some asteroids, found close to the ecliptic plane, have magnitudes fainter than 24 mag. These are potential candidates for small Main-Belt objects (e.g. icphg2010, j9bk75010). Other trails that are present at high geocentric ecliptic latitudes, have large differential motions, and pass relatively close to \textit{HST} are candidates for Near-Earth objects (NEOs; e.g. jdrz23010).

An example of a particular trail corresponding to a fast-moving object is in jds47w010. The curved long trail (based on the trail speed rate) may indicate a very close object. Another group of interest are objects with small differential rates and faint magnitudes. These could be distant objects, which are difficult to image from the ground (e.g. jcng06010).

\section{Discussion}
\label{discussionsection}

The use of archival data produced by imaging campaigns whose primary science goals lie outside the Solar System is common practice in asteroid science. There are several groups that used various image archives to find and characterise SSOs.  Some of the most relevant results were presented by \cite{Gwyn2012}, \cite{Vaduvescu2013}, \cite{Carry2016}, \cite{Popescu2016}, \cite{Vaduvescu2017}, \cite{Mahlke2018_SSO}, \cite{Mahlke2019}, \cite{Cortes2019}, \cite{Vaduvescu2020}, and \citet{Racero2022}.

A detailed description of the small bodies in the Solar System puts constraints on the different Solar System formation scenarios, which make concrete predictions on the size and orbit distribution of objects as a function of time \citep{2015aste.book..701B}. In particular, both the giant planets migrations and collisional cascades have effects on the asteroids size and orbital distributions that could be detectable with specially purposed observational surveys \citep{Morbidelli2005,Morbidelli2009}. However, since such surveys are expensive and hard to realise due to competition for telescope time, we instead decided to produce such a survey from a large archival dataset. 

\subsection{Comparison to previous work}
\label{previous_work}

\citet{Evans1998} presented a first analysis of asteroid trails in \textit{HST} images. Using a visual identification method on consecutively taken raw \textit{HST} images, the authors found 230 images with moving objects in 28\,460 WFPC2 images, corresponding to 96 distinct objects. This was followed by an updated study, which found 263 additional images with trails (corresponding to 113 objects) in 75\,000 WFPC2 exposures taken until 2000 \citep{Evans2002}. Instead of using the individual \textit{HST} images, in this work we used composite \textit{HST} exposures. We find 1\,701 asteroid trails in 37\,323 composite images, which correspond to 118\,814 individual exposures (each composite images is made, on average, by 3.2 individual exposures). Therefore, we find that, on average, 1.4\% of individual \textit{HST} exposures contain trails. This is slightly higher than the 0.8\% and 0.4\% found by \citet{Evans1998} and \citet{Evans2002}, respectively. This difference can be explained by the different asteroid detection method used and the different instruments explored, ACS and WFC3 having increased sensitivity over WFPC2 and larger FoVs (202\arcsec and 160\arcsec vs. 150\arcsec). 

Previous works have used artificial intelligence methods to identify Solar Systems objects, either using ground-based observations \citep{Ivezic2001, Duev2019} or images obtained from space observatories, such as simulated images for the future Euclid mission \citep{Lieu2019}. These methods can also be used to distinguish potential future Earth-impacting objects in existing asteroid catalogues \citep{Hefele2020}. Additionally, citizen science projects has also been used to discover NEAs in existing surveys (e.g. in SDSS, \citealt{Solano2014}, or in the Catalina Sky Survey, \citealt{Beasley2013}).

In this work, we suggest an additional practical approach: combining citizen science and deep learning to mine entire instrument archives, taking advantage of cloud computing resources. This novel approach complements the unique challenges and capabilities of the \textit{HST} exposures: a variety of instrumental and observational parameters giving rise to a multitude of asteroid trail characteristics in the images. Asteroid trails in \textit{HST} images have been known and referenced for a long time due to their clear and apparent paths, even for faint objects, thanks to the unique \textit{HST} characteristics. Nevertheless, as outlined above, the recovery of these trails required a sophisticated combination of human and artificial intelligence. While the total number of recovered asteroid observations is a fraction compared to the efforts on ground-based wide-field surveys, the appeal in the \textit{HST} trails is based in the remarkably faint objects that are serendipitously captured.

\subsection{Further work}
\label{further_work}

The next step of our work will be to analyse the light curves from the trails to detect rotational properties. Typical \textit{HST} exposure times range from several minutes to several hours, which limits the detection of full light curves to fast rotating asteroids (longer period objects would be difficult to distinguish). The parallax-induced curvature of the trails in the images can be used to infer asteroid orbital parameters \citep{Evans1998}. This is another interesting axis of our ongoing work. Once the distance to the object is estimated, we can obtain its absolute magnitude and get an estimate of its size. This will be presented in a follow-up paper of this project.

Finally, our study presents results of citizen science and deep learning applied to detect asteroids in the eHST archive. An obvious follow-up task will be to use transfer learning to re-purpose our classification algorithm to be applied to other datasets, both from ESA missions and from other large astronomical datasets. In particular, this methodology can be applied to classify the multiple high-quality ground-based surveys. As our results suggest, the heterogeneity of \textit{HST} observations used for this project has not been an obstacle for our algorithm to detect asteroids, so we would probably expect a better performance on a survey-originated dataset.

\section{Conclusions}
\label{conclusionssection}

We have performed a first large-scale exploration of the \textit{HST} archive of images taken in the last 20 years for serendipitously observed trails of SSOs. For this, we built a citizen science project on the Zooniverse platform, Hubble Asteroid Hunter, and trained an automated classifier based on deep learning. We find that:

\begin{itemize}
    \item Asteroid trails appear predominantly as curved trails in the \textit{HST} images due to the parallax induced by the motion of \textit{HST} around the Earth.
    \item We find asteroid trails in 3.5\% of \textit{HST} composite images, with a mean exposure time of 35 minutes, and in 1.4\% of the individual \textit{HST} exposures. 
    \item We detected 1\,701 asteroid trails, out of which we identified 670 trails (39\%) with 454 known SSOs and did not find matches for 1\,031 (61\%) trails. These are probably new, yet to be identified, fainter objects. 
    \item Of the identified SSOs, 95\% are Main-Belt SSOs and only 5\% are other types (Hungarias, NEAs).
    \item The asteroids we detected are in the 18-25 magnitude range. Asteroids that were not matched with known SSOs are 1.6 magnitudes fainter, on average, compared to the known SSOs. This suggests that \textit{HST} is more sensitive to fainter objects than what is accessible from the ground.  
    \item Of the detected asteroids, 96\% are within 30$^{\circ}$ of the ecliptic and only 4\% are at higher ecliptic latitudes, which is suggestive of highly inclined orbits. 
    \item We find a sky density of 80 asteroids per square degree for asteroids with magnitudes <24.5 mag close to the ecliptic, which decreases to $\sim$1 asteroid per square degree at high ecliptic latitudes. 
    \item Our archival exploration also allowed us to label other interesting objects in the \textit{HST} observations, such as trails of artificial satellites or arcs of strong gravitational lenses. These will be explored in future work. 
\end{itemize}

Since it is not a survey designed to map the ecliptic, \textit{HST} provides a unique, unbiased, and long-time-baseline study of SSOs. In future work, we will further investigate the still-unidentified  1\,031 trails by using the parallax induced by \textit{HST}. By fitting the trail shapes for parallax, a unique feature of space-based observatories, we will determine the geometric distances to the new objects and determine the statistics distribution of their sizes. 

Our study demonstrates a novel use of archival data with the modern tools of citizen science and deep learning and shows their potential in data-mining future big data surveys, such as Euclid or LSST.

% Add conclusion

\begin{acknowledgements}
We acknowledge the tremendous work made by over 11,000 citizen scientist volunteers on the Hubble Asteroid Hunter citizen science project. Their contributions are individually acknowledged on \url{https://www.zooniverse.org/projects/sandorkruk/hubble-asteroid-hunter/about/results}. We thank the anonymous referee for their helpful comments that have improved the paper. We also thank Robin Evans for the useful discussions on their previous work on asteroid trail identification in the \textit{HST} archives. We acknowledge support from Mar\'{\i}a Ar\'evalo and Javier Espinosa, from the {\it eHST} team at ESAC, who supported this research by providing postage stamps of \textit{HST} exposures on scale from the archive. Finally, we acknowledge our frequent and helpful interactions with the STScI helpdesk. This publication uses data generated via the Zooniverse.org platform, development of which is funded by generous support, including a Global Impact Award from Google, and by a grant from the Alfred P. Sloan Foundation. SK gratefully acknowledges support from the European Space Agency (ESA) Research Fellowship. The work of MP was supported by a grant of the Romanian National Authority for  Scientific  Research -- UEFISCDI, project number PN-III-P1-1.1-TE-2019-1504. This material is based upon work supported by the Google Cloud Research Credits program with the award GCP19980904. This work has made extensive use of data from {\it HST} mission, hosted by the European Space Agency at the {\it eHST} archive, at ESAC (\url{https://www.cosmos.esa.int/hst}), thanks to a partnership with the Space Telescope Science Institute, in Baltimore, USA (\url{https://www.stsci.edu/}) and  with the Canadian Astronomical Data Centre, in Victoria, Canada (\url{https://www.cadc-ccda.hia-iha.nrc-cnrc.gc.ca/}).\\

\end{acknowledgements}

\bibliographystyle{aa} % style aa.bst
\bibliography{bibliography.bib} % your references Yourfile.bib

\clearpage
\begin{appendix}

% \clearpage
% \newpage
% \onecolumn
% \LTcapwidth=\textwidth

\onecolumn
\begin{landscape}
\section{Table with the asteroid trails identified in the Hubble Space Telescope observations}
\begin{table}
\caption{Example of the log corresponding to 20 trails identified in \textit{HST} images. The following information is shown: the \textit{HST} composite observation ID, the instrument, the filter used, the number of individual exposures in which the SSO was identified ($N.ex.$),  the start time of the exposure in which the SSO was first identified ($MJD_{start}$, provided in modified Julian days), the end time of the last exposure in which the SSO was identified ($MJD_{end}$, provided in modified Julian days), the total exposure time for the trail ($texp$) -- obtained as the sum of the exposure times of all individual exposures in which the trail was identified; it can be different from $MJD_{end}$ - $MJD_{start}$ because there were breaks between individual exposures for readout, filter changes etc. -- the right ascension (RA) and declination (Dec) for the beginning and for the end of the trail, the total length of the trail, the apparent magnitude in the filter used for the observation ($Mag.$), and the corresponding asteroid (if it was identified; otherwise, a question mark indicates unidentified objects).}
\label{FullTable}
\tiny
\begin{tabular}{c l l l c c c c c c c c c c r}
No. & \textit{HST} obs. id & Instrument & Filter & $N.ex.$ & $MJD_{start}$ & $MJD_{end}$ & $texp[sec]$ & $RA_{start}[^\circ]$ & $Dec_{start}[^\circ]$ & $RA_{end}[^\circ]$ & $Dec_{end}[^\circ]$ & $Length [$\arcsec$]$ & $Mag.$ & Asteroid  \\ \hline
1 & ib1901010 & WFC3/UVIS & F438W & 4 & 55275.47375081 & 55275.50736207 & 2520 & 146.7329823 & 10.0969372 & 146.72596 & 10.0971402 & 25.71 & 21.234 & 234839 \\
2 & ib2r03020 & WFC3/UVIS & F475W & 2 & 55059.89004167 & 55059.9070326 & 1340 & 57.8625451 & 28.3094616 & 57.866537 & 28.3105017 & 13.519 & 23.278 &  ? \\
3 & ib4801010 & WFC3/UVIS & F475W & 2 & 55328.67625205 & 55328.69824294 & 1772 & 135.3464886 & 18.2331854 & 135.3422318 & 18.2361665 & 18.774 & 23.487 & 2015 PJ1 \\
4 & ib4803010 & WFC3/UVIS & F475W & 4 & 55268.17325339 & 55268.25039429 & 3686 & 135.0326681 & 22.5615199 & 135.0202912 & 22.549794 & 61.16 & 23.023 &  ? \\
5 & ib4803020 & WFC3/UVIS & F814W & 3 & 55268.28160969 & 55268.31696837 & 2799 & 135.0172415 & 22.5440068 & 135.0042467 & 22.5320243 & 63.213 & 22.538 &  ? \\
6 & ib4a28010 & WFC3/UVIS & F475X & 2 & 55138.5709588 & 55138.58695413 & 720 & 35.9802863 & -6.8473467 & 35.9856543 & -6.8479701 & 19.986 & 23.795 &  ? \\
7 & ib4a28010 & WFC3/UVIS & F475X & 2 & 55138.5709588 & 55138.58695413 & 720 & 36.0062025 & -6.8745683 & 36.0007509 & -6.8762862 & 21.238 & 23.805 &  ? \\
8 & ib4a28010 & WFC3/UVIS & F475X & 2 & 55138.5709588 & 55138.58695413 & 720 & 35.9964364 & -6.8643625 & 36.0010168 & -6.8639096 & 16.906 & 24.193 &  ? \\
9 & ib4a28020 & WFC3/UVIS & F600LP & 2 & 55138.57687323 & 55138.5928686 & 720 & 35.9777738 & -6.8464616 & 35.9837244 & -6.8479207 & 22.864 & 23.239 &  ? \\
10 & ib4a28020 & WFC3/UVIS & F600LP & 2 & 55138.57687323 & 55138.5928686 & 720 & 36.0043732 & -6.8754162 & 35.9987301 & -6.8763896 & 20.802 & 23.558 &  ? \\
11 & ib4a28020 & WFC3/UVIS & F600LP & 2 & 55138.57687323 & 55138.5928686 & 720 & 35.994681 & -6.8641965 & 35.999243 & -6.8642717 & 16.737 & 23.630 &  ? \\
12 & ib5n44010 & WFC3/UVIS & F625W & 5 & 55371.13967909 & 55371.16736443 & 1880 & 152.2859317 & 7.2036321 & 152.2772089 & 7.2066865 & 33.716 & 20.600 & 27949 \\
13 & ib5t01020 & WFC3/UVIS & F475W & 2 & 55050.83167392 & 55050.85093295 & 1536 & 37.7796172 & -8.5240886 & 37.7821469 & -8.5260418 & 11.754 & 23.969 &  ? \\
14 & ib5t04020 & WFC3/UVIS & F475W & 4 & 55248.23207211 & 55248.25373299 & 1486 & 128.8535618 & 43.8547617 & 128.8468845 & 43.8569909 & 20.462 & 21.959 &  ? \\
15 & ib5t11020 & WFC3/UVIS & F475W & 2 & 55284.54470008 & 55284.56388985 & 1530 & 163.5283395 & 4.4755862 & 163.5235591 & 4.4789158 & 22.393 & 20.451 & 63689 \\
16 & ib5t15020 & WFC3/UVIS & F475W & 1 & 55050.00431281 & 55050.01316667 & 765 & 210.1252985 & -1.5180055 & 210.1225965 & -1.5160961 & 12.771 & 22.289 & 517042 \\
17 & ib6w37020 & WFC3/UVIS & F438W & 1 & 55335.59584722 & 55335.60024517 & 380 & 189.7940639 & -0.5086657 & 189.7932223 & -0.5091481 & 3.711 & 22.354 & 2014 EL7 \\
18 & ib7g05010 & WFC3/UVIS & F606W & 4 & 55250.51128507 & 55250.53327558 & 1516 & 180.8701654 & 18.0475822 & 180.8778802 & 18.0439606 & 31.444 & 21.494 &  ? \\
19 & ib7g05010 & WFC3/UVIS & F606W & 1 & 55250.52888912 & 55250.53327558 & 379 & 180.8792933 & 18.0758788 & 180.8832089 & 18.0763413 & 14.552 & 21.066 &  ? \\
20 & ib93d2010 & WFC3/UVIS & F606W & 3 & 55536.04347844 & 55536.05963603 & 1140 & 141.3955949 & 31.7434672 & 141.3962258 & 31.7418404 & 6.709 & 23.835 &  ? \\
… & … & … & … & … & … & … & … & … & … & … & … & … & … & … \\

\hline
\end{tabular}
\end{table}
\end{landscape}

\end{appendix}

\end{document}